\documentclass[prd,showpacs,showkeys,superscriptaddress,nofootinbib,twocolumn,floatfix]{revtex4}
\usepackage{amsfonts}
\usepackage{graphicx,color,amsmath,amssymb}
\usepackage{subfigure}
\usepackage[pdfborder={0 0 0}]{hyperref}

\def\blfootnote{\xdef\@thefnmark{}\@footnotetext}
\newcommand{\eg}{\textit{e.g.}}

\newcommand{\ie}{\textit{i.e.}}

\newcommand{\ma}{\mathrm}
\newcommand{\ml}{\mathcal}

\newcommand{\ud}{\ma{d}}
\newcommand{\prel}{p_{\ma{rel}}}
\newcommand{\vprel}{\vec{p}_{\ma{rel}}}

\begin{document}

\title{Simulation of heavy quarkonium equilibration in the quark-gluon plasma}
\author{Shouxing Zhao}
\email{sxzhao2018@njust.edu.cn}
\affiliation{Department of Applied Physics, Nanjing University of Science and Technology, Nanjing 210094, China}
\author{Min He}
\email{minhephys@gmail.com}
\affiliation{Department of Applied Physics, Nanjing University of Science and Technology, Nanjing 210094, China}

\date{\today}

\begin{abstract}
We simulate the heavy quarkonium equilibration through transport in a static and homogeneous quark-gluon plasma (QGP) box within the semi-classical Boltzmann approach incorporating both the leading-order and next-to-leading-order dissociation and regeneration reactions. The scattering amplitudes involved are taken from perturbative computations based on effective color-electric dipole coupling of the heavy quarkonium with thermal gluons. By coupling the Langevin simulation of single heavy quark diffusion and the Boltzmann transport of the heavy quarkonium in a real-time fashion, we demonstrate how the kinetic and chemical equilibrium of heavy quarkonium with single heavy quarks in the medium is achieved in terms of both the bound state's yields and momentum distributions. The pertinent equilibration time turns out to be comparable to the lifetime of the QGP created in the most central heavy-ion collisions at the LHC energies. The role of the intricate interplay between the open and hidden heavy sector in the process of equilibration is highlighted. This work provides a dynamical way of understanding the phenomenological success of statistical hadronization model for charmonium production in relativistic heavy-ion collisions, and also paves the way for realistic phenomenological applications to heavy quarkonium transport.
\end{abstract}


\maketitle

\section{Introduction}
\label{sect:intro}

Heavy quarkonia below the open heavy-meson threshold are considered to be non-relativistic bound states of charm ($c$) and bottom ($b$) quark-antiquark pairs, as primarily justified by the large quark masses $m_{c,b}\gg\Lambda_{\rm QCD}$. While the vacuum spectroscopy of these bound states represent a versatile laboratory for testing the pertinent binding force, the spectral and abundance modifications of heavy quarkonia
in the Quark-Gluon Plasma (QGP) environment as created in relativistic heavy-ion collisions provide useful probes of the color force and transport properties of the deconfined medium~\cite{Zhao:2020jqu, He:2022ywp, Andronic:2024oxz}. In this regard, the primary observation was the weakening of the in-medium heavy quark potential due to the Debye screening of the static color charge surrounded by liberated thermal partons, which eventually leads to the melting of bound states at sufficiently high temperatures~\cite{Matsui:1986dk,Kluberg:2009wc}. As a result, the heavy quarkonium yield in heavy-ion collisions would be suppressed compared to that produced in proton-proton ($pp$) collisions and scaled by the hard binary collision number, which has been corroborated in particular by the suppression pattern of $\Upsilon(nS)$ ($s$ wave $b{\bar b}$ bound states) mesons in the sequence of their vacuum binding energies~\cite{CMS:2023lfu}. It has been argued in the statistical hadronization model that color screening might melt all initially produced charmonia and therefore the final production of charmonia is simply given by their statistical equilibrium yields~\cite{Andronic:2017pug,Andronic:2021erx,Andronic:2025jbp}.

The heavy quarkonium in-medium dynamics turns out much richer than the pure static screening picture. First the transitions from bound states to unbound heavy quark pairs, {\it i.e.} dissociation caused by the dynamical scattering off the medium partons, can reduce the heavy quarkonium yield. The dissociation reactions include the $2\rightarrow2$ gluo-dissociation (also known as leading-order dissociation) $\Psi+g\rightarrow Q+\bar{Q}$ ($\Psi$ denotes a heavy quarkonium and $Q$($\bar{Q}$) the heavy quarks) and the $2\rightarrow3$ partonic inelastic scattering (also known as next-to-leading order dissociation) $\Psi+p\rightarrow Q+\bar{Q}+p$ with thermal partons $p=g,q/\bar{q}$. The former is analogous to the photo-dissociation of neutral atoms via a dipole transition and has long been studied since the seminal work by Peskin~\cite{Peskin:1979va,Bhanot:1979vb}. The latter, as first proposed in~\cite{Grandchamp:2001pf} from the ``quasifree" perspective, overcomes the phase-space mismatch inherent in the gluo-dissociation process when the incident gluon energy is significantly larger than the bound state binding energy. This $2\rightarrow3$ process has since also been intensively studied from dynamical scattering point of view~\cite{Song:2005yd,Hong:2018vgp,Chen:2018dqg,Yao:2018sgn,Zhao:2024gxt} and from the Landau damping (or imaginary potential) point of view~\cite{Laine:2006ns,Beraudo:2007ky,Brambilla:2013dpa}. Inverse to the dissociation process, quarkonium bound state can be regenerated via recombination of single $Q$ and ${\bar Q}$ near in phase space~\cite{Braun-Munzinger:2000csl,Thews:2000rj,Grandchamp:2003uw}, in particular in situations where there are multiple heavy quark pairs diffusing and thermalizing in the QGP, {\it e.g.,} charm production in Pb-Pb collisions at the LHC energy. The regeneration production of charmonia and $B_c$ mesons has been corroborated by the experimental observations of the significant enhancement of the yields of these bound states at low transverse momenta $p_T$ and the $J/\psi$ large elliptic flow~\cite{ALICE:2020pvw,ALICE:2023gco,CMS:2022sxl}, and the pertinent theoretical explanations~\cite{Zhou:2014kka,He:2021zej,Wu:2023djn,Zhao:2024jkz}.

A natural theoretical framework to simultaneously incorporate dissociation and regeneration dynamics and thus to simulate the heavy quarkonium transport in the QGP is the semi-classical Boltzmann equation~\cite{Polleri:2003kn}. Various forms of this approach has been implemented in literature, considering only the $2\leftrightarrow2$ $\Psi+g \leftrightarrow Q+\bar{Q}$ process~\cite{Zhou:2014kka,Yao:2017fuc}, including also the $2\leftrightarrow 3$ $\Psi+p \leftrightarrow Q+\bar{Q}+p$ reactions~\cite{Song:2012at,Du:2022uvj,Hong:2019ade,Yao:2020xzw}, or via the reduced rate equation~\cite{Du:2017qkv}. A more modern treatment of the heavy quarkonium transport is the open quantum system (OQS) approach ~\cite{Brambilla:2016wgg,Akamatsu:2020ypb,Yao:2021lus}. The OQS approach keeps track of the evolution of $Q{\bar Q}$ pair wave packet in a hot medium via the Lindblad equation and takes care of the quantum transitions between different eigenstates. However, solving the Lindblad equation is numerically expensive and currently it is limited to the simulation of the dynamics of only one correlated $Q{\bar Q}$ pair~\cite{Brambilla:2020qwo,Miura:2019ssi,Miura:2022arv}. In the quantum optical limit (relatively low temperature where quarkonia are still tightly bound), it was demonstrated that the semi-classical Boltzmann equation can be derived from the OQS approach~\cite{Yao:2018nmy,Yao:2021lus}.

Given the complex nature of these transport approaches (various theoretical approximations involved, interplay between the microscopic reactions and macroscopic evolutions, and coupling the heavy quarkonium system to the dynamically expanding QGP background, etc.), it is theoretically desirable to investigate some limiting cases of heavy quarkonium transport even under simplified conditions, in order to ensure that the transport implementations work properly and basic principles check out before embarking on real phenomenological applications. In this respect, it is principally important to theoretically delineate and numerically demonstrate how the heavy quarkonium kinetic and chemical equilibration occurs should the system be given long enough time to evolve, which is not only interesting on its own right, but may also unravel in a dynamical way the rationale underlying the phenomenological success of the statistical hadronization model for describing the production of charmonia~\cite{Andronic:2017pug,Andronic:2021erx,Andronic:2025jbp} and $B_c$ mesons~\cite{Zhao:2024jkz} in Pb-Pb collisions at the LHC energy. This is exactly the purpose of the present work.

The simulation of heavy quarkonium equilibration has been previously pursued using semiclassical Langevin~\cite{Young:2008he,Oei:2024zyx} or Boltzmann transport~\cite{Yao:2017fuc}, and the OQS approach~\cite{Miura:2022arv}. Distinct from these works that are limited to color-singlet attractive potential, $2\leftrightarrow2$ reactions, or one pair of correlated $Q{\bar Q}$, we simulate the heavy quarkonium equilibration via Boltzmann equation incorporating both the $2\leftrightarrow2$ $\Psi+g \leftrightarrow Q+\bar{Q}$ and $2\leftrightarrow3$ $\Psi+p \leftrightarrow Q+\bar{Q}+p$ reactions, with pertinent realistic scattering amplitudes computed from our previous work~\cite{Chen:2017jje,Zhao:2024gxt}. Limiting our consideration to a finite number of $Q{\bar Q}$ (more specifically, 10 pairs of $c$ and $\bar{c}$ quarks resembling the realistic case in central Pb-Pb collisions at the LHC energies) and one species of heavy quarkonium (specifically the ground state charmonium $J/\psi$) in a static and homogeneous QGP box, while coupling the dissociation/regeneration reactions of the $J/\psi$ with the Langevin diffusion of single $c$ quarks, we keep track of the time evolution of the yields as well as the momentum distributions for the bound state. We are able to demonstrate that, although lagging behind and dependent on the thermalization of single $c$ quark kinetics, the dissociation and regeneration dynamics always drives the $J/\psi$ yields toward {\it relative} chemical equilibrium with the open $c$ and $\bar{c}$ quarks in the box as dictated by the statistical balance equation~\cite{Andronic:2021erx}, and the $J/\psi$ momentum distributions also come to thermal equilibrium, irrespective of the initial conditions explored. In particular, catalysis-like $2\leftrightarrow3$ reactions are shown to accelerate the heavy quarkonium equilibration at high temperatures due to their larger reaction rates compared to the $2\leftrightarrow2$ processes~\cite{Hong:2018vgp,Zhao:2024gxt}. This study not only lays solid foundation for phenomenological applications, but also highlights the interplay between open and hidden charm transport, especially through the off-diagonal recombination of $c$ and $\bar{c}$ quarks forming bound state charmonium, thus reiterating the findings previously obtained in the context of addressing the $J/\psi$'s large elliptic flow~\cite{He:2021zej}.

This article is structured as follows. In the subsequent Sec.~\ref{sect:scattering}, the microscopic reactions that underlies the heavy quarkonium dissociation and regeneration dynamics are discussed, where the transition amplitudes serving as input for the transport equation are illustrated. In Sec.~\ref{sect:boltzmann}, we first spell out the equilibrium limit as dictated by the statistical balance between open $c$ and $\bar{c}$ quark and the charmonium bound state ($J/\psi$). Then we move on to discuss the formalism of the Boltzmann transport; in particular the implementation of dissociation and regeneration are discussed in detail in connection with the $2\leftrightarrow2$ and $2\leftrightarrow3$ reactions we have considered. Sec.~\ref{sect:simul} is devoted to the numerical simulations and results, where the heavy quarkonium equilibration is first simulated assuming fully thermalized $c$ and $\bar{c}$ quarks, which is then generalized to the more general initial heavy quark distributions by coupling the single $c$'s diffusion via Langevin simulation to the Boltzmann transport of $J/\psi$. For both cases, precise relative chemical equilibrium of $J/\psi$ with open $c$ and $\bar{c}$ quarks are shown to be finally well reached. We summarize in Sec.~\ref{sect:sum}.

\section{Chemical reactions of quarkonium in QGP}
\label{sect:scattering}
The chemical reactions of heavy quarkonium leading to the dissociation and regeneration in the QGP are discussed in this section, including leading order (LO, $2\leftrightarrow2$) and next-to-leading order (NLO, $2\leftrightarrow3$) processes. The LO reaction occurs through absorbing or inversely radiating a real thermal gluon, $\Psi+g\leftrightarrow Q+\bar{Q}$, which dominates at low temperatures when the binding energy is larger than the screening mass of the medium~\cite{Chen:2017jje}. At high temperatures the NLO inelastic scattering processes ($\Psi+p\leftrightarrow Q+\bar{Q}+p$ with $p=g,q/\bar{q}$) become important because of new phase space being opened up~\cite{Grandchamp:2001pf,Zhao:2024gxt}. The scattering amplitudes of these processes can be computed by identifying the transition matrix element from $Q\bar{Q}$ color singlet state $|S\rangle$ to an octet state $|O,a\rangle$ ($a$ denotes the color indices) based on the gauge-invariant color-electric dipole coupling of the compact quarkonium with external soft gluons ~\cite{Yan:1980uh, Brambilla:2004jw, Sumino:2014qpa}

\begin{align}\label{V_SO}
	\langle O,a|V_{SO}|S \rangle = \frac{g_s}{\sqrt{2N_c}}\vec{E}^a(t,\vec{x})\cdot \langle O|\vec{r}|S\rangle,
\end{align}
where $\vec{E}^a$ denotes the color-electric field, $r$ the relative distance between $Q$ and $\bar{Q}$ and $g_s$ the strong coupling constant.

Using the above transition matrix element supplemented by standard three-gluon or quark-gluon vertex, we have calculated the transition amplitudes for the LO and NLO dissociation processes within the framework of old-fashioned quantum-mechanical perturbation theory
\begin{align}\label{Tfi}
	T_{fi} &= \langle f|H_{\ma{int}}|i\rangle +\sum_{m} \frac{ \langle f|H_{\ma{int}}|m\rangle \langle m|H_{\ma{int}}|i\rangle}{E_i - E_m +i\epsilon} \nonumber\\
	&=T_{fi}^{\ma{LO}} + T_{fi}^{\ma{NLO}}.
\end{align}
for a heavy quarkonium sitting at rest in the QGP~\cite{Chen:2017jje,Chen:2018dqg,Zhao:2024gxt}, where $H_{\ma{int}}$ denotes the coupling Hamiltonian and $|i\rangle$, $|f\rangle$ and $|m\rangle$ the initial, final and intermediate states, and the box-normalization convention was used. In the present work, the heavy quarkonium transport in the QGP is simulated with Boltzmann equation that embodies the Lorentz-invariant transition amplitudes for the chemical reactions; therefore the above amplitudes $T_{fi}$ with box-normalization needs to be converted into the relativistically normalized (Lorentz-invariant) form $\ml{M}$ via~\cite{Weinberg:1995mt,Schwartz-QFT:2014}
\begin{align}
	\label{transform}
	|\ml{M}|^2=|T_{fi}|^2 \frac{1}{(2\pi)^3 \delta^3(\vec{p}_i - \vec{p}_f ) V } \prod_{i}(2E_iV) \prod_{f}(2E_fV),
\end{align}
where $V$ is the volume used in the box-normalization.

For the ground state charmonium $J/\psi$ ($|nlm\rangle=|100\rangle$) under consideration, the Lorentz invariant amplitude squared (with the degeneracies of incident gluon and outgoing $Q\bar{Q}$ octet summed over) for the LO reaction ($\Psi+g \leftrightarrow Q+\bar{Q}$) as converted from $T_{fi}^{\text{LO}}$~\cite{Chen:2017jje} reads
\begin{align}\label{M_LO_sqd}
	\sum{|\mathcal{M}^{\text{LO}}|^2}=&d_g\frac{g_s^2\pi}{9}\omega_{\vec{k}}\Big[\int r^3\ud r j_1(\prel r) R_{10}(r) \Big]^2 \nonumber\\&\times(2 E_{\Psi})(2\omega_{\vec{k}})(2E_{Q})(2E_{\bar{Q}}) ,
\end{align}
where the gluon degeneracy is $d_g=2\cdot8=16$, the incident gluon energy $\omega_{\vec{k}}$ and the relative momentum between $Q$ and $\bar{Q}$ in the color octet $\prel=|\vprel|$. The term in the bracket in Eq.~(\ref{M_LO_sqd}) represents the transition matrix element between the bound state and unbound $Q\bar{Q}$ pair, in which $j_1$ is spherical Bessel function of the first order and $R_{10}(r)$ the radial wave function of $J/\psi$ computed by solving Schr{\"o}dinger equation with a temperature dependent screening potential~\cite{Karsch:1987pv,Chen:2017jje}.

The same quantity (again with the degeneracies of incident and outgoing gluons and outgoing $Q\bar{Q}$ octet summed over) for the NLO reaction involving gluons  ($\Psi+g \leftrightarrow Q+\bar{Q}+g$) as converted from $T_{fi}^{\text{NLO,g}}$~\cite{Zhao:2024gxt} is
\begin{align}\label{M_NLOg_sqd}
	\sum{|\mathcal{M}^{\text{NLO,g}}|^2}= &d_g\frac{g_s^4\pi}{ \omega_{\vec{k}} \omega_{\vec{\kappa}} \prel^2} \Big[ \int r^3\ud rj_1(\prel r) R_{10}(r) \Big]^2 \nonumber\\ &\Big[ \frac{ \omega_{\vec{\kappa}}-\omega_{\vec{k}} } {\omega_{(\vec{\kappa}-\vec{k})}^2 - (\omega_{\vec{\kappa}}-\omega_{\vec{k}})^2 }  \Big]^2  g(\vec{\kappa},\vprel,\vec{k})\nonumber\\
	&\times (2 E_{\Psi})(2\omega_{\vec{k}})(2E_{Q})(2E_{\bar{Q}})(2\omega_{\vec{\kappa}}),
\end{align}
where $\omega_{\vec{\kappa}}$ and $\omega_{(\vec{\kappa}-\vec{k})}$ are energies of the outgoing and the exchanged gluons respectively, and $ g(\vec{\kappa},\vprel,\vec{k})$ is a polynomial function of three momenta arising from summing over gluon polarizations (see Ref.~\cite{Zhao:2024gxt} for details). Similarly the Lorentz invariant amplitude squared (with the degeneracies of incident and outgoing light quarks/antiquarks and outgoing $Q\bar{Q}$ octet summed over) for the NLO reaction involving light quarks/antiquarks ($\Psi + q/\bar{q} \leftrightarrow Q+\bar{Q}+q/\bar{q}$) is given by
\begin{align}\label{M_NLOq_sqd}
	\sum{|\mathcal{M}^{\text{NLO,q}}|^2}= &d_q \frac{4g_s^4\pi} {9E_{\vec{k}} E_{\vec{\kappa}} \prel^2 } \Big[\int r^3 \ud rj_1(\prel r)R_{10}(r)\Big]^2\nonumber\\ &\Big[\frac{E_{\vec{\kappa}}-E_{\vec{k}}}{\omega_{(\vec{\kappa}-\vec{k})}^2-(E_{\vec{\kappa}}-E_{\vec{k}})^2}\Big]^2  f(\vec{\kappa},\vprel,\vec{k})\nonumber\\
	&\times (2 E_{\Psi})(2E_{\vec{k}})(2E_{Q})(2E_{\bar{Q}})(2E_{\vec{\kappa}}).
\end{align}
where $d_q=2\cdot2\cdot3\cdot3=36$, $E_{\vec{k}}$ and $E_{\vec{\kappa}}$ are energies of incident and outgoing light quarks/antiquarks, and the polynomial $f(\vec{\kappa},\vprel,\vec{k})$ comes from summation over light quark spins (see Ref.~\cite{Zhao:2024gxt} for details).

Although what we have computed are the amplitudes for the dissociation processes~\cite{Chen:2017jje,Zhao:2024gxt}, the amplitudes for the inverse regeneration reactions remain the same as guaranteed by the time reversal symmetry (detailed balance principle). We also note that in terms of the power counting of the effective field theory (EFT)~\cite{Brambilla:2004jw,Yao:2018sgn}, the LO reaction occurs at the order of $O(g_sr)$, while the NLO reactions are at the order of $O(g_s^2r)$~\cite{Zhao:2024gxt}. Apart from the dissociation and regeneration reactions, the quantum transitions between different quarkonium eigenstates, \eg, $J/\psi+g\leftrightarrow\psi(2S)+g$, have also been computed, which, however, appear at higher order of $O(g_s^2r^2)$ (involving the vertex of $V_{SO}$ (Eq.~(\ref{V_SO})) twice) and give significantly smaller rates than the dissociation/regeneration processes~\cite{Zhao:2024gxt}. We therefore neglect these higher-order transition processes in the present work.

\section{Semi-classical quarkonium transport with the Boltzmann equation}
\label{sect:boltzmann}

\subsection{Heavy quarkonium equilibrium limit}
\label{subsect:equil_limit}
In this work we consider a finite number of single $Q$ and $\bar{Q}$ pairs and one species of quarkonium (ground state $J/\psi$) that simultaneously transport in a static and homogeneous QGP box. We simulate the dissociation and regeneration reactions as discussed in the previous section and examine the equilibration for the bound state $\Psi$ in the QGP. Should the system be given sufficiently long time to evolve, one expects {\it relative} chemical equilibrium to be reached between open $Q\bar{Q}$ and the bound state $\Psi$, which is dictated by the balance equation in the statistical hadronization model (SHM)~\cite{Andronic:2017pug,Andronic:2021erx,Andronic:2025jbp}
\begin{align} \label{balanceEq}
	N_{Q\bar{Q}} = N^{\rm open}_{Q} + N_{\Psi}.
\end{align}
Eq.~(\ref{balanceEq}) expresses the ``balanced" partition of heavy quarks between open $Q\bar{Q}$ and the bound state quarkonium $\Psi$ in the equilibrium limit, under the constraint of heavy flavor number conservation. For a QGP box of fixed volume $V$ and homogeneous temperature $T$ that is much lower than the heavy quark mass,
\begin{align} \label{weight1}
	N^{\rm open}_{Q}&= d_{Q} \gamma_{Q} V \int \frac{\ud^3p_{Q}}{(2\pi)^3} e^{-E_{Q}(\vec{p}_{Q})/T} \nonumber\\
	&= d_{Q} \gamma_{Q} V \frac{m_Q^2T}{2\pi^2}K_{2}\left(\frac{m_Q}{T}\right),
\end{align}
and similarly
\begin{align} \label{weight2}
	N_{\Psi}&= d_{\Psi} \gamma_{Q}^2 V \frac{m_{\Psi}^2T}{2\pi^2}K_{2}\left(\frac{m_{\Psi}}{T}\right),
\end{align}
where $d_{Q}$ and $d_{\Psi}$ are the degeneracies of single $Q$ and the bound state $\Psi$, respectively, and the relativistic dispersion relation $E=\sqrt{m^2+\vec{p}^2}$ has been used to obtain $K_2$, the modified Bessel function of the second kind. The heavy quark fugacity factor, $\gamma_Q$, is introduced in Eqs.~(\ref{weight1}) and (\ref{weight2}), to characterize the deviation of realistic number of $Q\bar{Q}$ pairs in the system from the absolute chemical equilibrium limit at the given temperature~\cite{Andronic:2017pug,Andronic:2021erx,Andronic:2025jbp}, which should be selfconsistently determined from Eq.~(\ref{balanceEq}).

Throughout the work, we fix the number of $Q\bar{Q}$ pairs at $N_{Q\bar{Q}}=10$, which resembles the realistic situation of $c\bar{c}$ in central Pb-Pb collisions at the LHC energy~\cite{He:2021zej}. As shown by the horizontal dashed lines in Fig.~\ref{fig_input_thermalized_Q} and Fig.~\ref{fig_with_Langevin} in Sec.~\ref{sect:simul}, the computed equilibrium limit (with temperature dependent $c$-quark and $J/\psi$ masses~\cite{Zhao:2024gxt}) of the quarkonium yield increases as temperature decreases; that is, $Q\bar{Q}$'s tend to be existing in the form of bound states (with lower energies) at lower temperatures. To compare with pertinent existing studies in literature that also employed semi-classical approaches but obtained nonrelativistic equilibrium limit~\cite{Yao:2017fuc, Yao:2020xzw, Oei:2024zyx}, we have also computed such equilibrium limits with $E = m + \frac{\vec{p}^2}{2m}$ in Eqs.~(\ref{weight1}) and (\ref{weight2}).

\subsection{Boltzmann equation for quarkonium evolution}
\label{subsect:boltzmann}

The single-particle phase space distribution function for single $Q$ ($\bar Q$) or quarkonium $\Psi$ reads
\begin{align}
	f_{Q,\bar{Q},\Psi}(\vec{x},\vec{p},t) = \frac{\ud N_{Q,\bar{Q},\Psi}(\vec{x},\vec{p},t)} {\frac{d^3xd^3p}{(2\pi)^3} },
\end{align}
where the heavy quark fugacity factor and degeneracies have been implicitly incorporated. The $2\leftrightarrow2$ and $2\leftrightarrow3$ reactions elaborated on in Sec.~\ref{sect:scattering} are simulated by the Boltzmann equation that keeps track of the time evolution of the quarkonium distribution function
\begin{align}~\label{Boltzmann}
	&\frac{\ud}{\ud t} f_{\Psi}(\vec{x},\vec{p},t)=(\frac{\partial}{\partial t}+\frac{\partial \vec{x}}{\partial t}\cdot\nabla_{\vec{x}})f_{\Psi}(\vec{x},\vec{p},t)\nonumber\\
	&=C_{22}(\Psi+g \leftrightarrow Q + \bar{Q})+C_{23}(\Psi +p \leftrightarrow Q + \bar{Q} +p ).
\end{align}
Labeling the momenta involved in LO process as $\Psi(\vec{p}),~g(\vec{p}_2),~Q(\vec{p}_3),~\bar{Q}(\vec{p}_4)$ in the static QGP global frame, the $2 \leftrightarrow 2$ collision term in Eq.~(\ref{Boltzmann}), in terms of the invariant amplitude $\ml{M}^{\rm LO}$ in Eq.~(\ref{M_LO_sqd}), reads,
\begin{align}  \label{Bol_eq_LO}
	&C_{22}=C_{22}[{\rm gain}] - C_{22}[{\rm loss}]\nonumber\\
	&=\frac{1}{2E_{\Psi}(\vec{p})} \int \frac{\ud^3p_2}{2E_{g}(2\pi)^3} \frac{\ud^3p_3}{2E_{Q}(2\pi)^3} \frac{\ud^3p_4}{2E_{\bar{Q}}(2\pi)^3}\nonumber\\ &\sum|\ml{M}^{\text{LO}} |^2 (2\pi)^4 \delta^4(p+p_2-p_3-p_4)\nonumber\\
	&\times[\frac{d_{\Psi}}{d_{Q}d_{\bar{Q}}} f_{Q}(\vec{x},\vec{p}_3,t) f_{\bar{Q}}(\vec{x},\vec{p}_4,t) \left(1+f_{g}(\vec{x},\vec{p}_2,t)\right) \nonumber\\ &- f_{\Psi}(\vec{x},\vec{p},t) f_{g}(\vec{x},\vec{p}_2,t)],
\end{align}
Denoting the pertinent momenta $\Psi(\vec{p})$, $p(\vec{p}_2)$, $Q(\vec{p}_3)$, $\bar{Q}(\vec{p}_4)$, $p(\vec{p}_5)$ again in the QGP global frame and using the amplitudes $\ml{M}^{\rm NLO}$ in Eqs.~(\ref{M_NLOg_sqd}) and (\ref{M_NLOq_sqd}), the $2 \leftrightarrow 3$ collision term in Eq.~(\ref{Boltzmann}) can be similarly written as
\begin{align}  \label{Bol_eq_NLO}
	&C_{23}= C_{23}[{\rm gain}] - C_{23}[{\rm loss}]\nonumber\\
	&=\frac{1}{2E_{\Psi}(\vec{p})} \int \frac{\ud^3p_2}{2E_{p}(\vec{p}_2)(2\pi)^3} \frac{\ud^3p_3}{2E_{Q}(2\pi)^3} \frac{\ud^3p_4}{2E_{\bar{Q}}(2\pi)^3}  \nonumber\\ & \frac{\ud^3p_5}{2E_{p}(\vec{p}_5)(2\pi)^3} \sum|\ml{M}^{\text{NLO}} |^2 (2\pi)^4 \delta^4(p+p_2-p_3-p_4-p_5)\nonumber\\
	&\times[\frac{d_{\Psi}}{d_{Q}d_{\bar{Q}}} f_{Q}(\vec{x},\vec{p}_3,t) f_{\bar{Q}}(\vec{x},\vec{p}_4,t) f_{p}(\vec{x},\vec{p}_5,t) \left(1\pm f_{p}(\vec{x},\vec{p}_2,t)\right) \nonumber\\ &- f_{\Psi}(\vec{x},\vec{p},t) f_{p}(\vec{x},\vec{p}_2,t) \left(1\pm f_{p}(\vec{x},\vec{p}_5,t)\right) ].
\end{align}
In Eqs.~(\ref{Bol_eq_LO}) and (\ref{Bol_eq_NLO}), the Bose/Fermi distribution has been used for gluons/light quarks and the Bose-enhancement/ Pauli-blocking taken into account for final state gluons/light quarks with effective thermal parton masses~\cite{Zhao:2024gxt}. Since the amplitudes in Eqs.~(\ref{M_LO_sqd}), (\ref{M_NLOg_sqd}) and (\ref{M_NLOq_sqd}) are expressed in terms of the $\Psi$ rest frame variables, $|\ml{M}^{\text{LO}} |^2$ and $|\ml{M}^{\text{NLO}} |^2$ in Eqs.~(\ref{Bol_eq_LO}) and (\ref{Bol_eq_NLO}) should be evaluated by boosting the QGP frame momenta and energies to the rest frame of $\Psi$.

In the limit when heavy quarks are fully thermalized the above Boltzmann equation can be reduced to the rate equation for the phase space distribution of $\Psi$. To derive this, one notes that the dissociation rate at LO has already been implied in Eq.~(\ref{Bol_eq_LO}),
\begin{align}    \label{diss_rate_LO}
	\Gamma^{\text{LO}}&(p,T(\vec{x}))=\frac{1}{2E_{\Psi}(\vec{p})} \int \frac{\ud^3p_2}{2E_{g}(2\pi)^3} \frac{\ud^3p_3}{2E_{Q}(2\pi)^3} \frac{\ud^3p_4}{2E_{\bar{Q}}(2\pi)^3} \nonumber\\
	& \sum|\ml{M}^{\text{LO}} |^2 (2\pi)^4 \delta^4(p+p_2-p_3-p_4) f_{g}(\vec{x},\vec{p}_2,t).
\end{align}
On the other hand, for equilibrium phase distributions $f_{Q}^{\text{eq}}$ and $f_{\bar{Q}}^{\text{eq}}$ for the single heavy quarks one has
\begin{align}	\label{rate_eq_LO_condition}
	&\frac{d_{\Psi}}{d_{Q}d_{\bar{Q}}} f_{Q}^{\text{eq}}(\vec{p}_3) f_{\bar{Q}}^{\text{eq}}(\vec{p}_4) \left(1+f_{g}(\vec{p}_2)\right) \nonumber\\
	=&d_{\Psi} \gamma_{Q} e^{-E_{Q}(\vec{p}_3)/T} \gamma_{\bar{Q}} e^{-E_{\bar{Q}}(\vec{p}_4)/T} \left(1+\frac{1}{e^{E_{g}(\vec{p}_2)/T}-1}\right) \nonumber\\
	=&d_{\Psi}  \gamma_{Q}^2 e^{\left( -E_{Q}(\vec{p}_3)-E_{\bar{Q}}(\vec{p}_4)+E_{g}(\vec{p}_2) \right)/T} \frac{1}{e^{E_{g}(\vec{p}_2)/T}-1} \nonumber\\
	=&d_{\Psi}  \gamma_{Q}^2 e^{-E_{\Psi}(\vec{p})/T}  \frac{1}{e^{E_{g}(\vec{p}_2)/T}-1}\nonumber\\
	=& f_{\Psi}^{\text{eq}}(\vec{p}) f_g(\vec{p}_2),
\end{align}
where the $\vec{x}$ dependence has been omitted, and $f_{\Psi}^{\text{eq}}$ is the equilibrium phase space distribution for the bound state $\Psi$. Combining (\ref{diss_rate_LO}) and (\ref{rate_eq_LO_condition}), Eq.~(\ref{Bol_eq_LO}) can be simplified to the rate equation for the LO reaction
\begin{align}
	\label{rate_eq_LO}
	\frac{\ud}{\ud t}f_{\Psi}(\vec{x},\vec{p},t) = -\Gamma^{\text{LO}}(p,T(\vec{x})) \left[ f_{\Psi}(\vec{x},\vec{p},t) - f_{\Psi}^{\text{eq}}(\vec{x},\vec{p}) \right] .
\end{align}
Similarly, combining the NLO dissociation rate embodied in Eq.~(\ref{Bol_eq_NLO}),
\begin{align}	\label{diss_rate_NLO}
	&\Gamma^{\text{NLO,g/q}}(p,T(\vec{x}))= \frac{1}{2E_{\Psi}(\vec{p})} \int \frac{\ud^3p_2}{2E_{p}(2\pi)^3} \frac{\ud^3p_3}{2E_{Q}(2\pi)^3} \frac{\ud^3p_4}{2E_{\bar{Q}}(2\pi)^3}   \nonumber\\ & \frac{\ud^3p_5}{2E_{p}(2\pi)^3} \sum|\ml{M}^{\text{NLO}} |^2(2\pi)^4 \delta^4(p+p_2-p_3-p_4-p_5) \nonumber\\
	&\times f_{p}(\vec{x},\vec{p}_2,t) \left(1\pm f_{p}(\vec{x},\vec{p}_5,t)\right),
\end{align}
and
\begin{align}  \label{rate_eq_NLO_condition}
	&\frac{d_{\Psi}}{d_{Q}d_{\bar{Q}}} f_{Q}^{\text{eq}}(\vec{p}_3) f_{\bar{Q}}^{\text{eq}}(\vec{p}_4) f_{p}(\vec{p}_5) \left(1\pm f_{p}(\vec{p}_2)\right)  \nonumber\\
	=&d_{\Psi} \gamma_{Q}^2 e^{-\left( E_{Q}(\vec{p}_3) + E_{\bar{Q}}(\vec{p}_4) -E_{p}(\vec{p}_2)  \right)/T} \frac{1}{e^{E_{p}(\vec{p}_5)/T} \mp 1}  \frac{1}{e^{E_{p}(\vec{p}_2)/T} \mp 1} \nonumber\\
	=&d_{\Psi} \gamma_{Q}^2 e^{-\left( E_{\Psi}(\vec{p}) -E_{g}(\vec{p}_5) \right)/T} \frac{1}{e^{E_{p}(\vec{p}_5)/T} \mp 1}  \frac{1}{e^{E_{p}(\vec{p}_2)/T} \mp 1} \nonumber\\
	=& f_{\Psi}^{\text{eq}}(\vec{p})f_{p}(\vec{p}_2) \left(1\pm f_{p}(\vec{p}_5) \right),
\end{align}
Eq.~(\ref{Bol_eq_NLO}) is reduced to rate equation for NLO reaction
\begin{align}
	\label{rate_eq_NLO}
	\frac{\ud}{\ud t}f_{\Psi}(\vec{x},\vec{p},t) = -\Gamma^{\text{NLO,g/q}}(p,T(\vec{x})) \left[ f_{\Psi}(\vec{x},\vec{p},t) - f_{\Psi}^{\text{eq}}(\vec{x},\vec{p}) \right] .
\end{align}
Furthermore, if the dissociation rates involved in Eqs.~(\ref{rate_eq_LO}) and (\ref{rate_eq_NLO}) are replaced by the momentum-averaged ones, the phase space distributions can be integrated to produce the rate equation for the yield of $\Psi$
\begin{align}
	\label{rate_eq}
	\frac{\ud}{\ud t}N_{\Psi}(t) = -\Gamma(\langle p\rangle,T) \left[ N_{\Psi}(t) - N_{\Psi}^{\text{eq}} \right],
\end{align}
which has been widely used for describing the heavy quarkonium transport in literature~\cite{Grandchamp:2003uw,Rapp:2008tf}. In Eq.~(\ref{rate_eq}), the equilibrium yield $N_{\Psi}^{\text{eq}}$ of $\Psi$ now acts as a transport coefficient associated with regeneration. The above derivation of the rate equations suggests that heavy quarkonium equilibration through dissociation and regeneration dynamics relies on the kinetic thermalization of single $Q$ and $\bar{Q}$ in advance.

\subsection{Dissociation of quarkonium}
\label{subsect:disso}

\begin{figure} [!t]
	\includegraphics[width=\columnwidth]{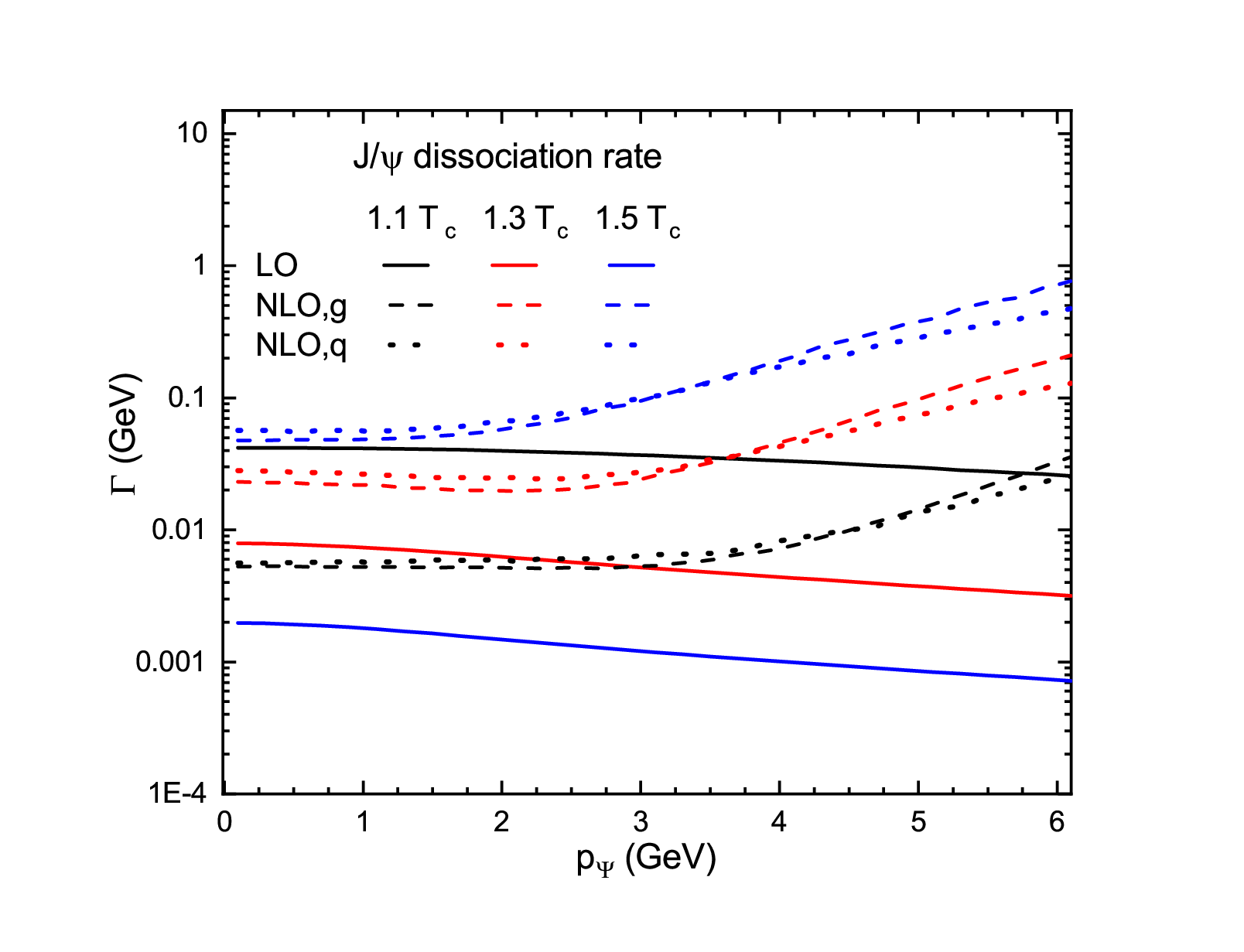}
	\includegraphics[width=\columnwidth]{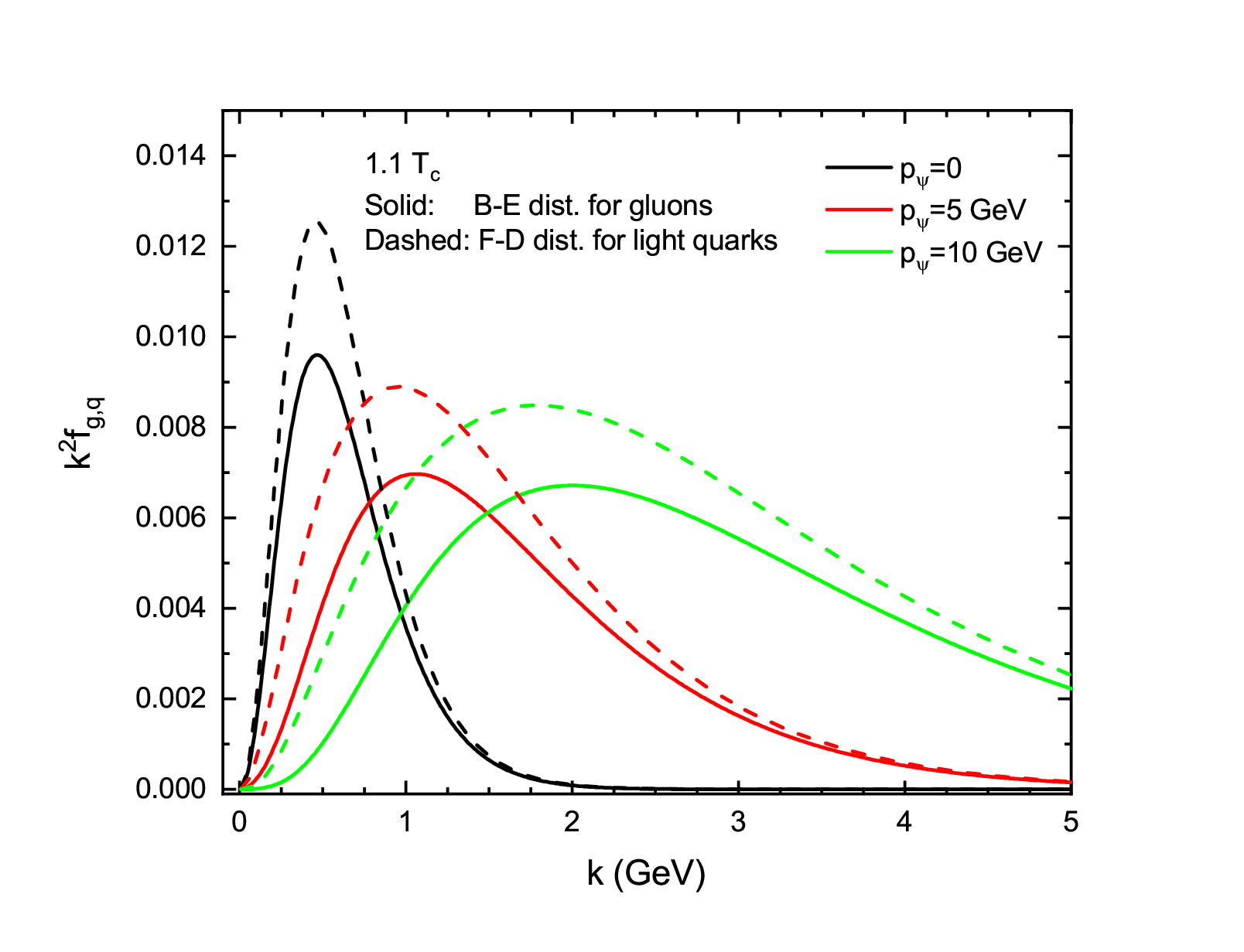}
	\caption{Upper panel: dissociation rates for $J/\psi$ with finite momentum at different temperatures; contributions from LO (solid), gluon-induced NLO (dashed) and light quark-induced NLO (dotted) processes are displayed separately. Lower panel: thermal distributions of incident partons seen by the heavy quarkonium as function of parton momentum, for varying $J/\psi$ momenta at 1.1$T_c$. }
	\label{fig_diss_rates}
\end{figure}

We first evaluate the dissociation rates as defined in Eqs.~(\ref{diss_rate_LO}) and (\ref{diss_rate_NLO}) for a heavy quarkonium moving in QGP, which will be used for the numerical simulations in Sec.~\ref{sect:simul}. By solving Schr{\"o}dinger equation with temperature dependent heavy quark potential, the dissociation temperature for the ground state charmonium $J/\psi$ has been determined to be at about $1.6T_c$ ($T_c$ being the pseudo-critical temperature)~\cite{Zhao:2024gxt}. The dissociation temperature is also considered to be the onset point for the bound state regeneration. Therefore in this work we perform the heavy quarkonium transport simulations in a static QGP at three constant temperatures: $T$=1.1$T_c$, 1.3$T_c$, 1.5$T_c$ for the purpose of illustration.

The numerical results for the dissociation rates $\Gamma(p,T)$ as function of $J/\psi$ momentum at different temperatures are displayed in the upper panel of Fig.~\ref{fig_diss_rates}. At the lowest temperature 1.1$T_c$, the LO dissociation rates dominate over the NLO counterparts in the low momentum region. But as temperature increases, the LO rates start to lose steam, while the NLO rates keep growing and take over. As for the momentum dependence of these rates, the LO rates monotonously decrease with increasing momentum but the NLO rates generally increase toward high momenta. The similar momentum dependence for the $\Psi$'s dissociation rates was also found in~\cite{Du:2017qkv,Hong:2019ade}. To understand these behaviors we examine the parton distribution function from the perspective of the rest frame of the $\Psi$. The thermal distribution of gluons or light quarks seen by the $\Psi$ reads $k^2/(e^{\gamma(E(\vec{k})+\vec{k}\cdot\vec{v})/T}\mp1)$, with $k$ being the thermal parton's momentum and $\gamma$ the Lorentz factor associated with the quarkonium's moving velocity $v$ in the QGP global frame. As shown in the lower panel of Fig.~\ref{fig_diss_rates}, for increasing momentum of the $\Psi$, the parton distributions shift towards higher momentum. Therefore, the convolution of LO cross section with the gluon distribution, \ie, LO dissociation rate, reduces with $\Psi$ momentum increasing, since the LO cross section for a quarkonium at rest peaks at relatively low incident gluon energy and falls off towards higher energy~\cite{Brambilla:2011sg,Chen:2017jje,Hong:2018vgp}. In contrast, the NLO cross section gradually grows with incident parton energy and finally saturates~\cite{Zhao:2024gxt}, which makes its overlap with the parton distribution function grow with the incident parton energy (measured in the $\Psi$'s rest frame), thereby yielding increasing rates for larger  $\Psi$ momentum as seen from the QGP global frame. We also note that the rates arising from the gluon-induced NLO process are slightly smaller at low momenta and become larger towards higher momenta compared to the light quark counterparts.

Taking advantage of the homogeneity of the static QGP box, one needs to only keep track of the $\Psi$'s momentum distribution $f_{\Psi}(\vec{p},t)$ obtained by integrating out the coordinates of the phase space distribution function. Using the dissociation rates $\Gamma(p,T)$, the iteration scheme for the loss in $f_{\Psi}(\vec{p},t)$ can be constructed as
\begin{align}  \label{diss_simul}
	\frac{\ud}{\ud t}f_{\Psi}(\vec{p},t)[\text{loss}] &= -\int \frac{\ud^3x}{(2\pi)^3} \Gamma(p,T) f_{\Psi}(\vec{x},\vec{p},t)\nonumber\\
	&= -\Gamma(p,T) f_{\Psi}(\vec{p},t).
\end{align}

\subsection{Recombination from single $Q$ and $\bar{Q}$}
\label{subsect:recom}
We move on to discuss the implementation of the gain terms of Boltzmann equation that corresponds to the regeneration from diffusing $Q$ and $\bar{Q}$. Generally the phase space distributions for single $Q$ and $\bar{Q}$ can be represented via test particle method
\begin{align} \label{distributions}
	f_{Q}(t,\vec{x},\vec{p}) = \sum_{n_Q=1}^{N_{Q}} (2\pi)^3 \delta^3(\vec{x}-\vec{x}_{n_Q}(t)) \delta^3(\vec{p}-\vec{p}_{n_Q}(t)) , \nonumber\\
	f_{\bar{Q}}(t,\vec{x},\vec{p}) = \sum_{n_{\bar{Q}}=1}^{N_{\bar{Q}}} (2\pi)^3 \delta^3(\vec{x}-\vec{x}_{n_{\bar{Q}}}(t)) \delta^3(\vec{p}-\vec{p}_{n_{\bar{Q}}}(t)).
\end{align}

First the gain term for $2\leftrightarrow2$ process can be cast into a more tractable form in terms of the single $Q$ and $\bar{Q}$ test particles, by substituting Eq.~(\ref{distributions}) into Eq.~(\ref{Bol_eq_LO}) and integrating over the coordinate,
\begin{align} \label{gain_LO}
	&\frac{\ud}{\ud t}f_{\Psi}(\vec{p},t)[\text{gain},2\leftrightarrow2] = \int \frac{\ud^3x}{(2\pi)^3} C_{22}[\text{gain}]\nonumber\\
	&=\frac{d_{\Psi}}{d_{Q}d_{\bar{Q}}} \frac{1}{2E_{\Psi}(\vec{p})} \sum_{n_Q=1}^{N_{Q}} \sum_{n_{\bar{Q}}=1}^{N_{\bar{Q}}} \int \frac{\ud^3p_2}{2E_{g}(2\pi)^3} \frac{1}{2E_{Q}}  \frac{1}{2E_{\bar{Q}}} \nonumber\\
	&\sum|\ml{M}^{\text{LO}} |^2 (2\pi)^4 \delta^4(p+p_2-p_{n_Q}-p_{n_{\bar{Q}}}) \nonumber\\
	&\times \int \frac{\ud^3x}{(2\pi)^3} \delta^3(\vec{x}-\vec{x}_{n_Q}(t)) \delta^3(\vec{x}-\vec{x}_{n_{\bar{Q}}}(t)) \left(1+f_{g}(\vec{x},\vec{p}_2,t)\right)  \nonumber\\
	&=\frac{d_{\Psi}}{d_{Q}d_{\bar{Q}}} \frac{1}{2E_{\Psi}(\vec{p})} \sum_{n_Q=1}^{N_{Q}} \sum_{n_{\bar{Q}}=1}^{N_{\bar{Q}}} \frac{1}{2E_{g}} \frac{1}{2E_{Q}} \frac{1}{2E_{\bar{Q}}}  \sum|\ml{M}^{\text{LO}} |^2 \nonumber\\
	&(2\pi) \delta( E_{\Psi}(\vec{p}) + E_g(\vec{p}_2) - E_{Q}(\vec{p}_{n_Q}) - E_{\bar{Q}}(\vec{p}_{n_{\bar{Q}}}) ) \nonumber\\
	&\times \frac{1}{(2\pi)^3}\delta^3(\vec{x}_{n_Q}(t)-\vec{x}_{n_{\bar{Q}}}(t))  \left(1+f_{g}(\vec{x},\vec{p}_2,t)\right) ,
\end{align}
where in the last step, $\vec{p}_2 = -\vec{p} + \vec{p}_{n_Q} + \vec{p}_{n_{\bar{Q}}} $, and the gluon position is taken to be $\vec{x} = \big(\vec{x}_{n_Q}(t)+\vec{x}_{n_{\bar{Q}}}(t)\big)/2$.
In practice the delta function involving the spatial separation between $Q$ and $\bar{Q}$ is smeared by a Gaussian function
\begin{align}
	\delta^3 ( \vec{x}_{n_Q}(t) - \vec{x}_{n_{\bar{Q}}}(t) ) \approx  \frac {\exp[-(\vec{x}_{n_Q}(t)-\vec{x}_{n_{\bar{Q}}}(t))^2/(2\sigma^2)]} {(2\pi\sigma^2)^{3/2}}.
\end{align}
It has been numerically checked that, the choice of the spatial scale $\sigma$ in a relatively large range, \eg, $\sigma\in[0.15~\ma{fm}, 0.75~\ma{fm}]$, does not affect the numerical results. This is because, when a smaller (larger) $\sigma$ is employed, the effective number of $Q\bar{Q}$ pairs that are encompassed and participate in recombination is reduced (increased), which is, however, balanced by the larger (smaller) recombination probability arising from the enhanced (reduced) strength of the sharper (wider) Gaussian peak. As a result, the final results are rather robust against the variation of $\sigma$, when averaged over a large number of events in the simulation. Practically, since $\sigma$ characterizes the recombination range in coordinate space, one can choose $\sigma$ to be equal to the Bohr radius of bound state $\Psi$~\cite{Yao:2017fuc,Yao:2020xzw}. Finally, for relativistic $Q$ and $\bar{Q}$, the $\delta$ function ensuring energy conservation in Eq.~(\ref{gain_LO}): $f(p_z) = \sqrt{m_{\Psi}^2 +\vec{p}_T^2 + p_z^2} + \sqrt{m_g^2 + \vec{p}_{2,T}^2 + p_{2,z}^2} - \sqrt{m_Q^2+\vec{p}_{n_Q}^2} - \sqrt{m_{\bar{Q}}^2+\vec{p}_{n_{\bar{Q}}}^2} = 0$, can be eliminated via
\begin{align}
	\frac{\ud N}{p_T\ud p_T \ud \phi_p \ud p_z} \bigg|_{f(p_z)=0} \propto \frac{\int \ud p_z \delta(f(p_z))}{\Delta p_z},
\end{align}
which becomes accurate when $\Delta p_z$ is small enough.

Next by plugging Eq.~(\ref{distributions}) into Eq.~(\ref{Bol_eq_NLO}) and integrating out the coordinates, the gain term due to the $2\leftrightarrow3$ process can be similarly handled
\begin{align} \label{gain_NLO}
	&\frac{\ud}{\ud t}f_{\Psi}(\vec{p},t)[\text{gain},2\leftrightarrow3] = \int \frac{\ud^3x}{(2\pi)^3} C_{23}[\text{gain}]\nonumber\\
	&= \frac{d_{\Psi}}{d_{Q}d_{\bar{Q}}} \frac{1}{2E_{\Psi}(\vec{p})} \sum_{n_Q=1}^{N_{Q}} \sum_{n_{\bar{Q}}=1}^{N_{\bar{Q}}} \int \frac{\ud^3p_2}{2E_{p}(\vec{p}_2)(2\pi)^3} \frac{1}{2E_{Q}} \frac{1}{2E_{\bar{Q}}} \nonumber\\
	&\frac{\ud^3p_5}{2E_{p}(\vec{p}_5)(2\pi)^3}  \sum|\ml{M}^{\text{NLO}} |^2  (2\pi)^4 \delta^4(p+p_2-p_{n_Q} - p_{n_{\bar{Q}}} -p_5)\nonumber\\
	&\times\frac{1}{(2\pi)^3}\delta^3(\vec{x}_{n_Q}(t)-\vec{x}_{n_{\bar{Q}}}(t))  f_{p}(\vec{x},\vec{p}_5,t) \left(1\pm f_{p}(\vec{x},\vec{p}_2,t)\right) \nonumber\\
	&= \frac{d_{\Psi}}{d_{Q}d_{\bar{Q}}} \frac{1}{2E_{\Psi}(\vec{p})} \sum_{n_Q=1}^{N_{Q}} \sum_{n_{\bar{Q}}=1}^{N_{\bar{Q}}} \int \frac{p_2^2\ud\Omega_{\vec{p}_2}}{2E_{p}(\vec{p}_2)(2\pi)^3} \frac{1}{2E_{Q}} \frac{1}{2E_{\bar{Q}}} \nonumber\\
	&\frac{1}{2E_{p}(\vec{p}_5)} \sum | \ml{M}^{\text{NLO}} |^2  \frac{2\pi}{|f'(p_2)|} \frac{1}{(2\pi)^3} \delta^3(\vec{x}_{n_Q}(t) - \vec{x}_{n_{\bar{Q}}}(t)) \nonumber\\ &\times  f_{p}(\vec{x},\vec{p}_5,t) \left(1\pm f_{p}(\vec{x},\vec{p}_2,t)\right),
\end{align}
where $\vec{p}_5 = \vec{p}+\vec{p}_2-\vec{p}_{n_Q}-\vec{p}_{n_{\bar{Q}}}$, and $p_2=|\vec{p}_2|$ is calculated from $f(p_2) = E_{\Psi}(\vec{p})+E_p(\vec{p}_2) - E_Q(\vec{p}_{n_Q}) - E_{\bar{Q}}(\vec{p}_{n_{\bar{Q}}}) - E_p(\vec{p}_5)=0$, leaving only integration over the polar angle of $\vec{p}_2$.

In summary, Eqs.~(\ref{gain_LO}) and (\ref{gain_NLO}) represent the iteration scheme for the gain (due to the LO and NLO regeneration/recombination reactions) in the ${\Psi}$'s momentum distribution $f_{\Psi}(\vec{p},t)$, which, in combination with the iteration scheme constructed for the loss in $f_{\Psi}(\vec{p},t)$ (cf. Eq.~(\ref{diss_simul})), constitute the core method underlying our simulation for the ${\Psi}$'s equilibration. We also note that, the regeneration dynamics of heavy quarkonium $\Psi$ in the present approach is driven by the realistic quantum mechanical scattering amplitudes involving the participant $Q$, $\bar{Q}$ and light partons, in which the bound state wave function has been explicitly incorporated, cf. Eqs.~(\ref{M_LO_sqd}-\ref{M_NLOq_sqd}). This is in marked contrast to a recent study~\cite{Oei:2024zyx} on the same topic based on measuring the relative energy of the $Q$ and $\bar{Q}$ pairs. For the latter, an attractive force term arising from the real part of the complex heavy quark potential~\cite{Beraudo:2007ky} was added on top of the Langevin equation for the $Q$ and $\bar{Q}$'s diffusion. A bound state was then considered to be formed (destructed) as long as the relative energy between $Q$ and $\bar{Q}$ is negative (positive), which amounts to a purely classical criterion involving no quantum wave function effects.
Our method is similar to the one used in~\cite{Yao:2017fuc,Yao:2020xzw} where the dynamical scattering amplitudes were also employed to describe the transition between bound state and unbound heavy quark pairs. But in these latter works the simulation was performed in a full Monte-Carlo way such that the $\Psi$'s were also treated as test particles by sampling the differential rates and the equilibrium limit obtained was non-relativistic~\cite{Yao:2017fuc,Yao:2020xzw}.

\section{Numerical Simulations and Results}
\label{sect:simul}

As specified before, our simulations for quarkonium equilibration are performed in a static QGP box of fixed volume $L^3=(10~\text{fm})^3$ and homogeneous temperature $T$. The initial positions of heavy quarks are uniformly sampled and the periodic boundary condition is imposed for test particles $Q$ and $\bar{Q}$ in a finite size box. Because of the latter, when performing the recombination of $Q$ and $\bar{Q}$, the relative distance within that pair is counted by
\begin{align}\label{minimum-image-convention}
	r^i = \text{min}\left\{ |x^{i}_{Q} - x^{i}_{\bar{Q}}|, L-|x^{i}_{Q} - x^{i}_{\bar{Q}}| \right\},~ i=1,2,3,
\end{align}
with $x^{i}_{Q/\bar{Q}}$ being the real-time position of $Q/\bar{Q}$ in $i$th direction. Eq.~(\ref{minimum-image-convention}) is known as minimum image convention and widely adopted in literature~\cite{Miura:2019ssi}.

\subsection{Thermalized $Q$ and $\bar{Q}$}
\label{subsect:thermal_QQbar}

\begin{figure} [!t]
	\includegraphics[width=\columnwidth]{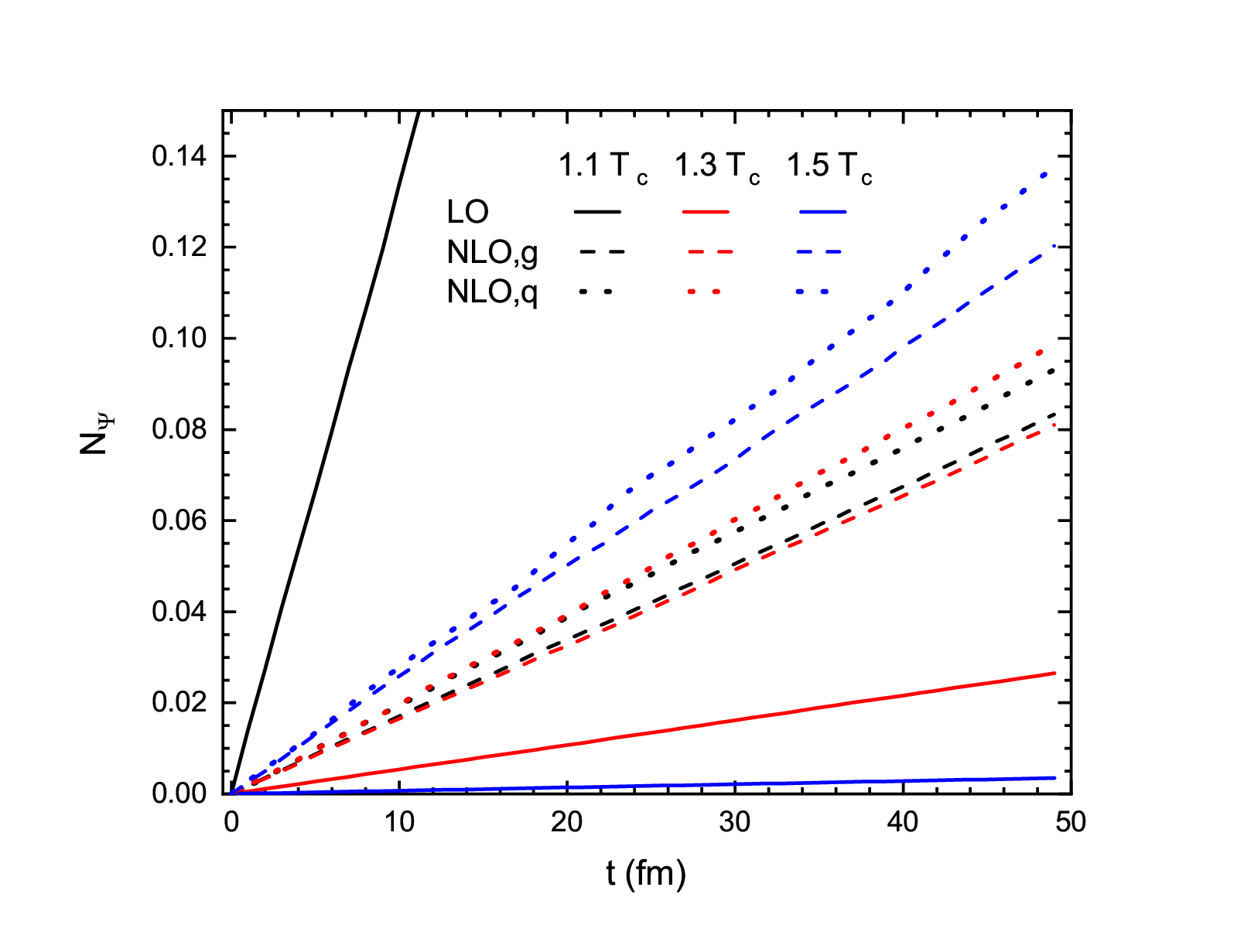}
	\includegraphics[width=\columnwidth]{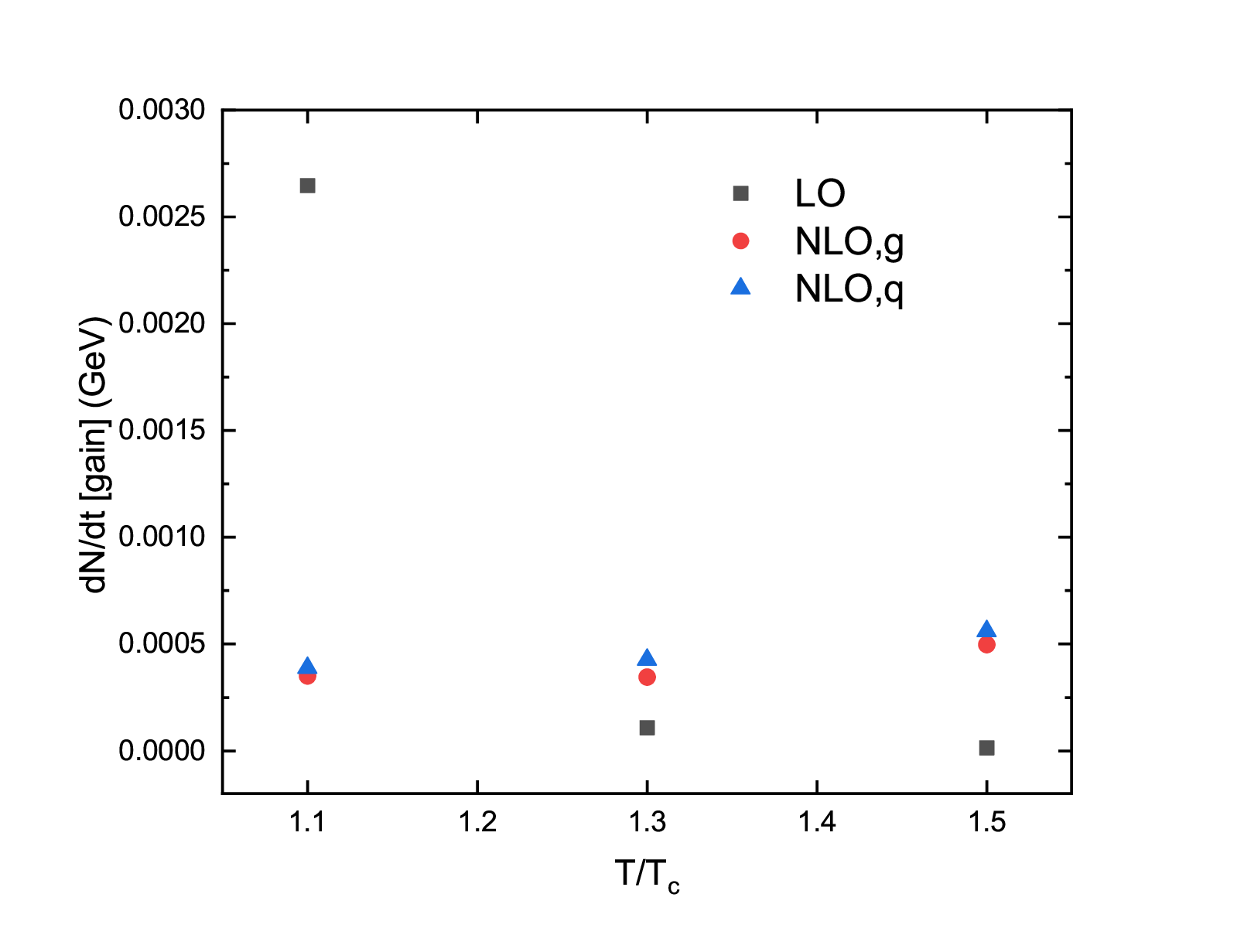}
	\includegraphics[width=\columnwidth]{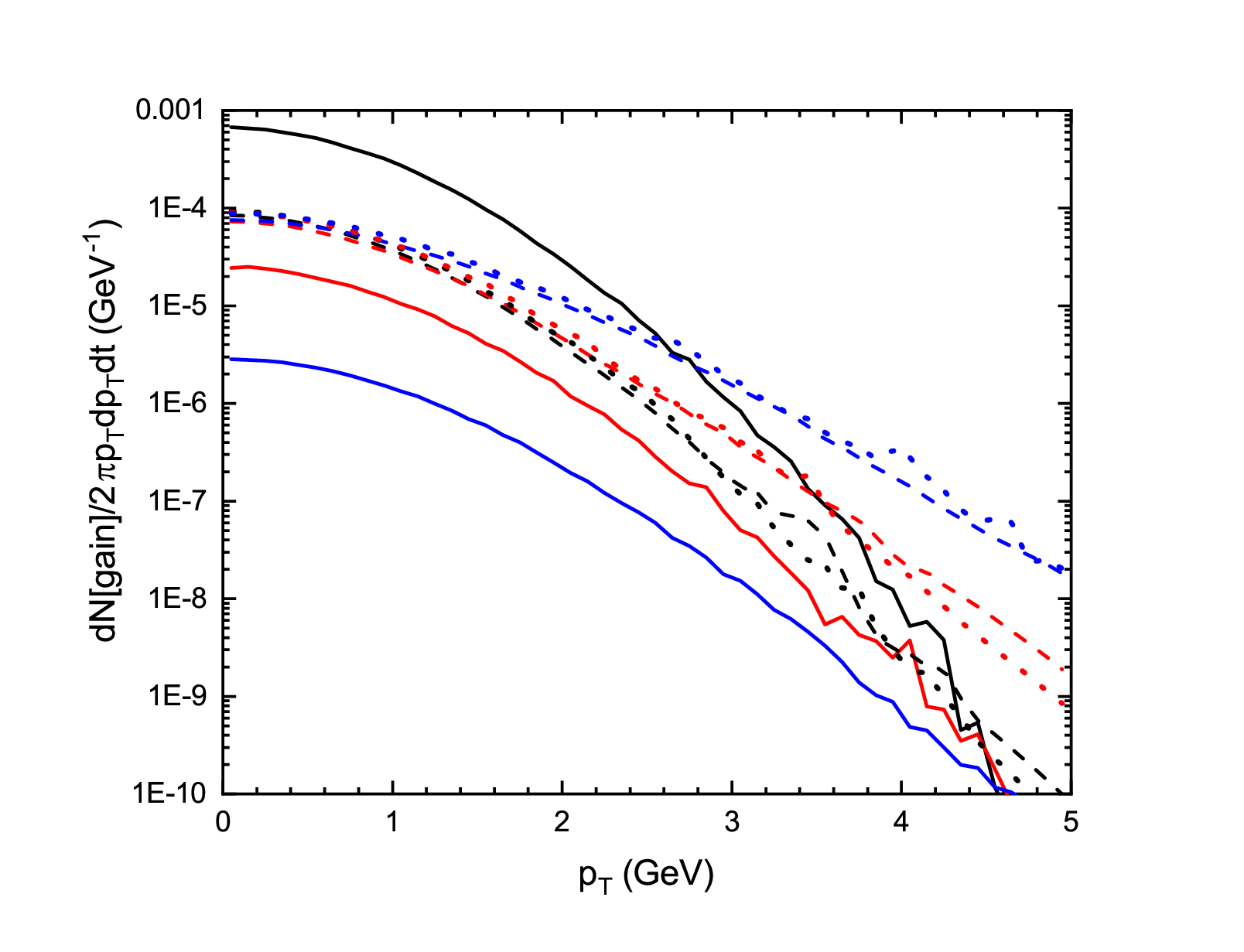}
	\caption{Upper panel: the time evolution of $J/\psi$ regenerated from 10 pairs of thermalized $c$ and $\bar{c}$ at varying temperatures; contributions from different dynamical processes are displayed separately.  Middle panel: regeneration rates extracted from upper panel. Lower panel: $p_T$-differential regeneration rates in the simulations.}
	\label{fig_gain_rates}
\end{figure}

We first perform a test simulation for heavy quarkonium chemical equilibration in the extreme case of full thermalization of single heavy quarks before embarking on more realistic simulation with non-equilibrium (\ie, transported) heavy quarks.

Since the regeneration plays a key role for quarkonium equilibration, let us first quantify the features of pure regeneration through LO and NLO processes via Eqs.~(\ref{gain_LO}) and (\ref{gain_NLO}), by using 10 pairs of $c$ and $\bar{c}$ that are fully thermalized, \ie, $f(\vec{p}_Q) \propto e^{-E_Q(\vec{p}_Q)/T} = e^{-\sqrt{m_Q^2+\vec{p}_Q^2}/T}$. The time evolution of the yield of purely regenerated $\Psi$ is displayed in the upper of Fig.~\ref{fig_gain_rates} for three temperatures. At a fixed temperature, the regenerated $\Psi$ yield simply keeps growing with time at a constant slope that can be extracted as the {\it absolute} regeneration rate and presented in middle panel of Fig.~\ref{fig_gain_rates}. The LO regeneration rates are pronounced at 1.1$T_c$ but decline rapidly towards higher temperatures. In contrast, the NLO regeneration rates show a general increase with $T$ and take over from LO at high temperatures; the light quark-induced NLO regeneration rates are always slightly larger than the gluon-induced one. Since here the $Q$ and $\bar{Q}$'s are assumed fully thermalized, these behaviors of the regeneration rates can be understood from the gain term of the rate equation for the integrated yield~(Eq.~(\ref{rate_eq})): $dN_{\Psi}(t)/dt~{\rm [gain]} = \Gamma(\langle p\rangle,T) N_{\Psi}^{\text{eq}}$, where the temperature dependence of the dissociation rates (decreasing/increasing with temperature for LO/NLO) as shown in the upper panel of Fig.~\ref{fig_diss_rates} is strengthened/mitigated by the equilibrium yield $N_{\Psi}^{\text{eq}}$ which reduces toward high temperatures. Furthermore, the $p_T$-differential regeneration rates are displayed in the lower panel of Fig.~\ref{fig_gain_rates}. As temperature increases, the $p_T$-spectra of both LO and NLO regeneration rates become harder, simply because the momentum distribution of thermalized $Q$ and $\bar{Q}$ participating in recombination are harder. Another observation is that the the $p_T$-spectrum of the LO regeneration rates is always softer than that from the NLO process, which is attributed to the softer momentum dependence of the LO dissociation rates than the NLO ones (upper panel of Fig.~\ref{fig_diss_rates}) embodied in the gain term of the (now momentum dependent) rate equations Eqs.~(\ref{rate_eq_LO}) and (\ref{rate_eq_NLO}).

\begin{figure} [htbp]
	\includegraphics[width=0.8\columnwidth]{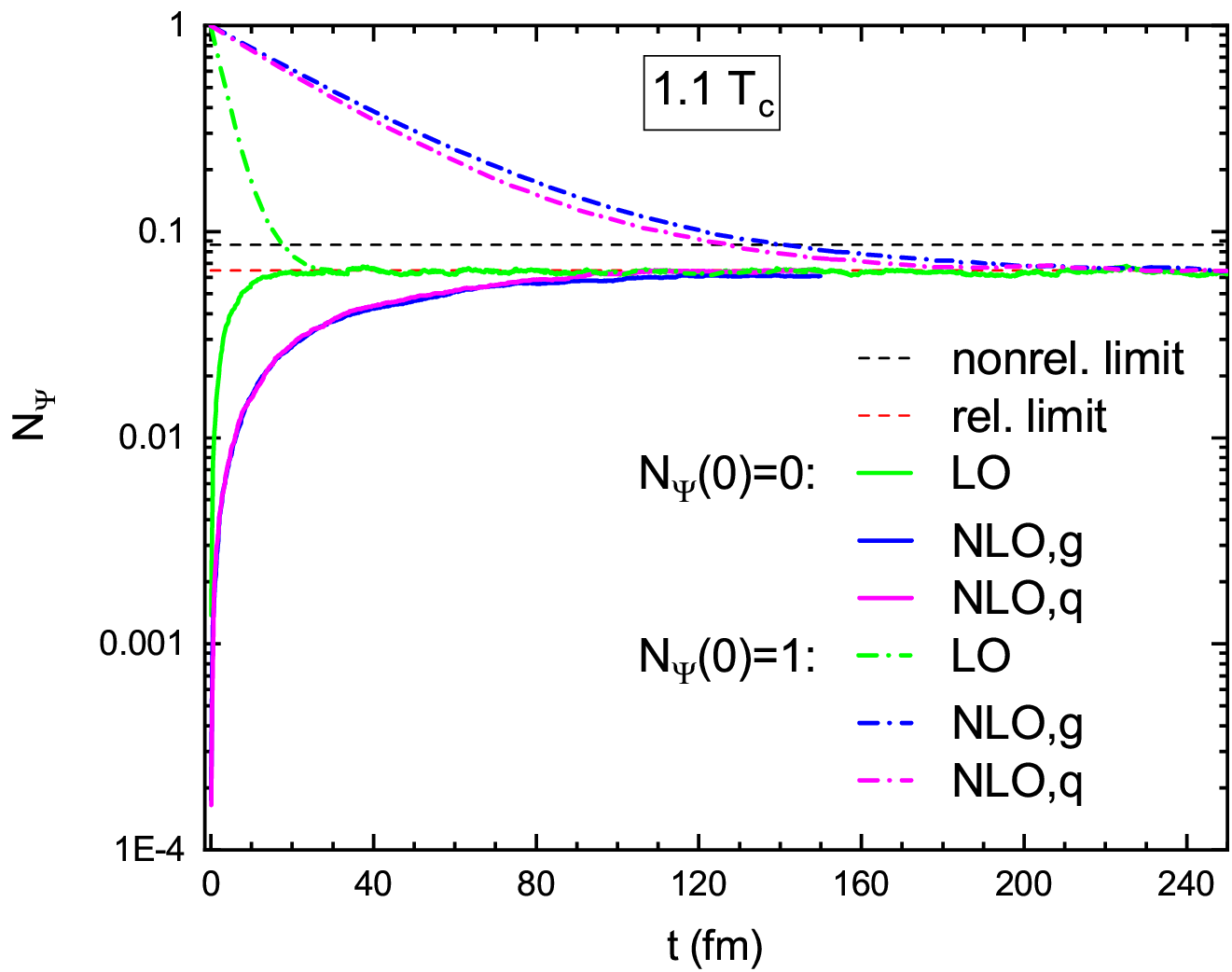}
	\includegraphics[width=0.8\columnwidth]{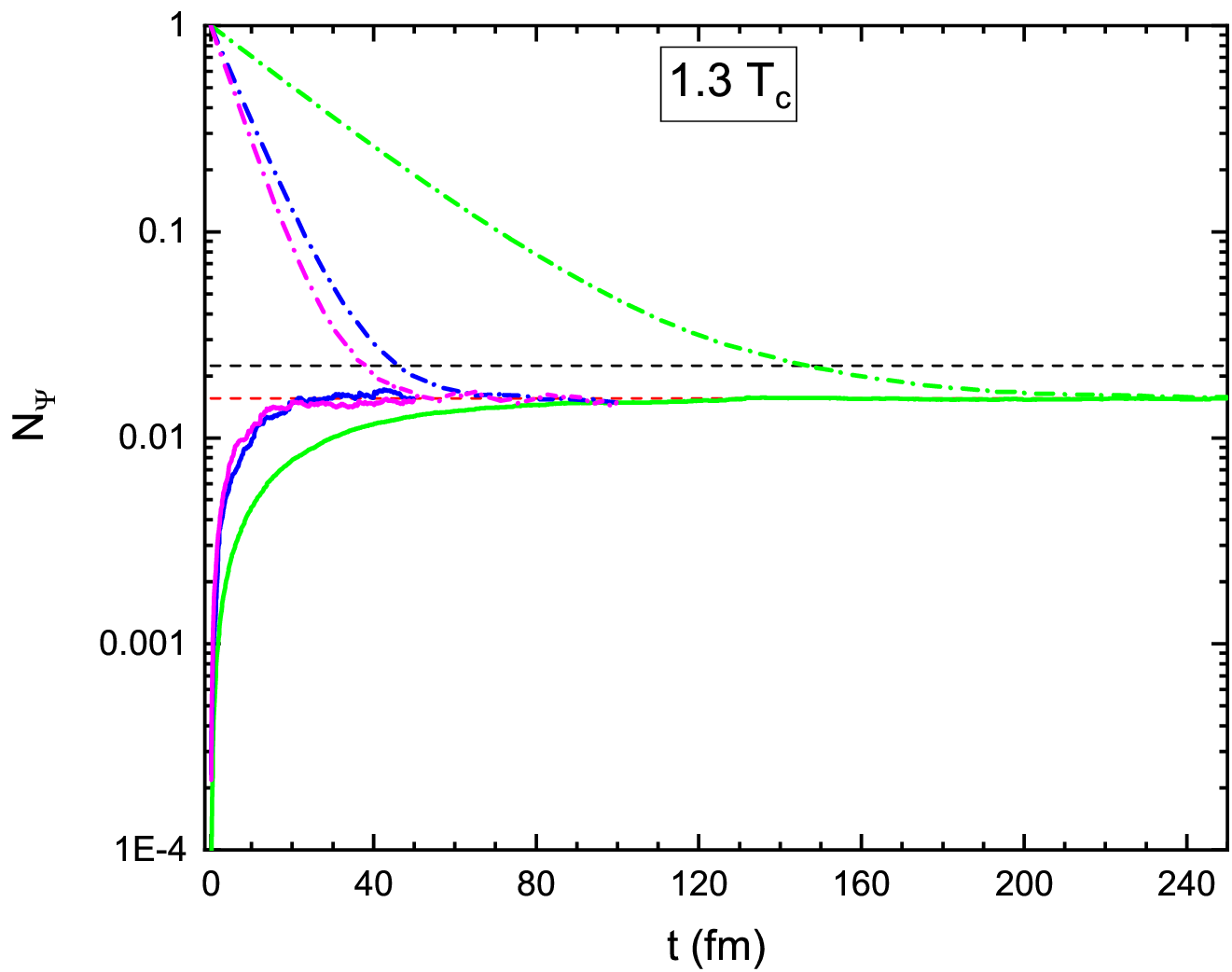}
	\includegraphics[width=0.8\columnwidth]{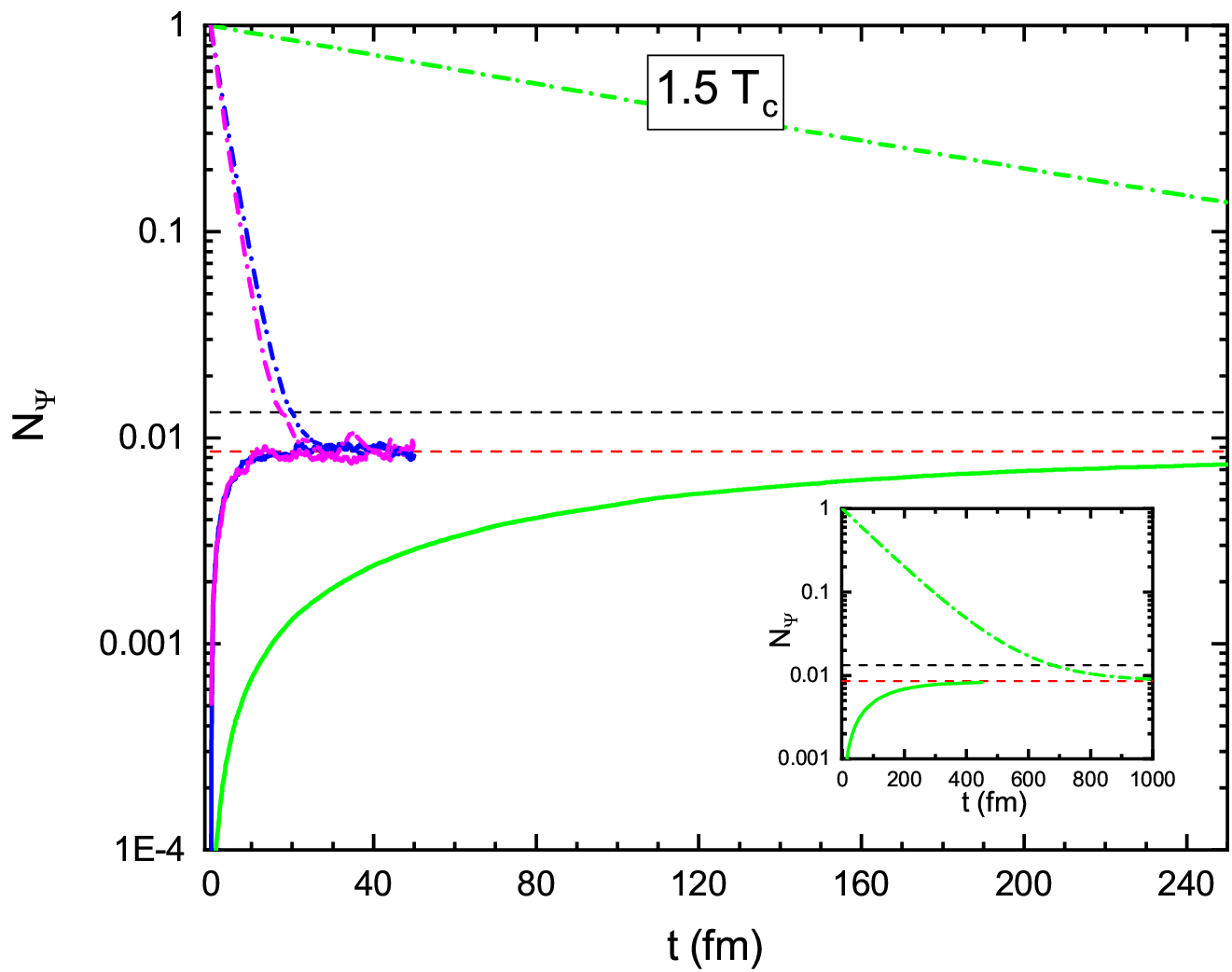}
	\caption{The time evolution of $J/\psi$ yield by simulating regeneration and dissociation simultaneously at 1.1$T_c$ (top), 1.3$T_c$ (middle) and 1.5$T_c$ (bottom), where the regeneration is computed from thermalized $c$ and $\bar{c}$. The horizontal black and red dashed lines correspond to the relative chemical equilibrium limits of $J/\psi$ integrated yields computed by the balance equation~(\ref{balanceEq}) with nonrelativistic and relativistic dispersion relations, respectively. Different initial conditions and different dynamical processes are shown for comparison.}
	\label{fig_input_thermalized_Q}
\end{figure}

Now we simulate the regeneration and the dissociation simultaneously. The conservation of heavy quarks, \ie, $N_{Q\bar{Q}}=N^{\rm open}_{Q}+N_{\Psi}=10$, is imposed throughout the evolution. Since the $Q$ and $\bar{Q}$'s participating in recombination are treated as test particles, their numbers are required to be integers. However as the evolution goes on, the real-time yield of the bound states is not an integer any more, neither is the number of $Q$ and $\bar{Q}$'s. To handle this situation, we envision that there are always $\ml{N}_{Q} = \ml{N}_{\bar{Q}} = 10$ and use them to compute the regeneration via Eqs.~(\ref{gain_LO}) and (\ref{gain_NLO}), but the resulting regenerated bound state yield should be rescaled by $(\ml{N}_{Q}-N_{\Psi}(t))(\ml{N}_{\bar{Q}}-N_{\Psi}(t)) / (\ml{N}_{Q}\ml{N}_{\bar{Q}})$, with $N_{\Psi}(t)$ being the real-time number of $\Psi$. This rescaling becomes more important when one has non-zero number of bound states at the initialization time.

We consider two kinds of initial conditions, \ie, initially there's no bound state ($N_{\Psi}(0)=0$) and initially there's one bound state($N_{\Psi}(0)=1$). For the latter case we set the $\Psi$ initial momentum distribution being thermalized. The numerical results for the time evolution of the integrated yield of $\Psi$ at three different temperatures are depicted in Fig.~\ref{fig_input_thermalized_Q}, where different dynamical processes (LO, NLO induced by gluons and NLO induced by light quarks/antiquarks) are compared. For both of the two initial conditions, after sufficiently long time evolution, the bound state abundance exhibits clear convergence toward the chemical equilibrium limit as given by the balance equation Eq.~(\ref{balanceEq}). We find that the time required for $\Psi$'s yield to reach equilibrium limit through LO reactions at 1.1$T_c$ is shorter compared to that through NLO processes, but at higher temperatures the LO chemical equilibration becomes much slower. For instance, at 1.3$T_c$, when there is no bound state initially, it takes $\sim30$ fm for NLO reactions to achieve chemical equilibrium, while the equilibration time for the LO reaction reaches $\sim80$ fm. This is in line with the fact that the LO rates (both the dissociation and regeneration, as discussed in Sec.~\ref{subsect:disso} and Sec.~\ref{subsect:recom}) are larger than NLO ones at low temperatures while the latter take over at high temperatures.
One also notices that at all temperatures and for all chemical reactions, when there's one bound state initially ($N_{\Psi}(0)=1$), the equilibration is always slowed down compared to there being no bound state initially ($N_{\Psi}(0)=0$), although the momentum distribution of initial bound state has been assumed to be thermal. For example still at 1.3$T_c$, the equilibration time with the initial condition $N_{\Psi}(0)=1$ reaches $\sim60$ fm and $\sim200$ fm for the NLO and LO reactions, respectively, which are around 2-2.5 times longer compared to the case with the initial condition $N_{\Psi}(0)=0$. This highlights the fact that the chemical equilibration always takes longer than the kinetic equilibration.

\subsection{Transported $Q$ and $\bar{Q}$}
\label{subsect:withLangevin}

\begin{figure} [htbp]
	\includegraphics[width=0.8\columnwidth]{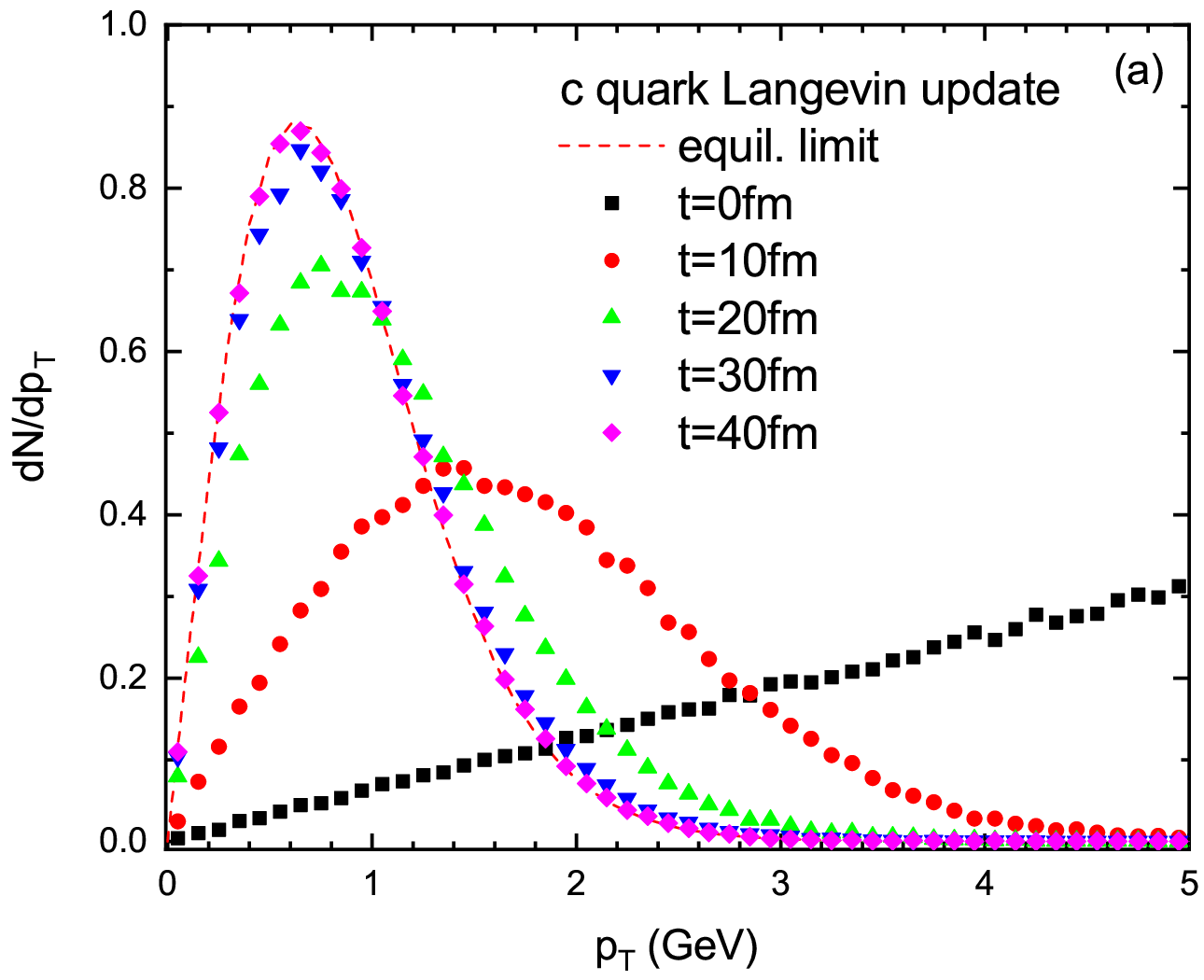}
	\includegraphics[width=0.8\columnwidth]{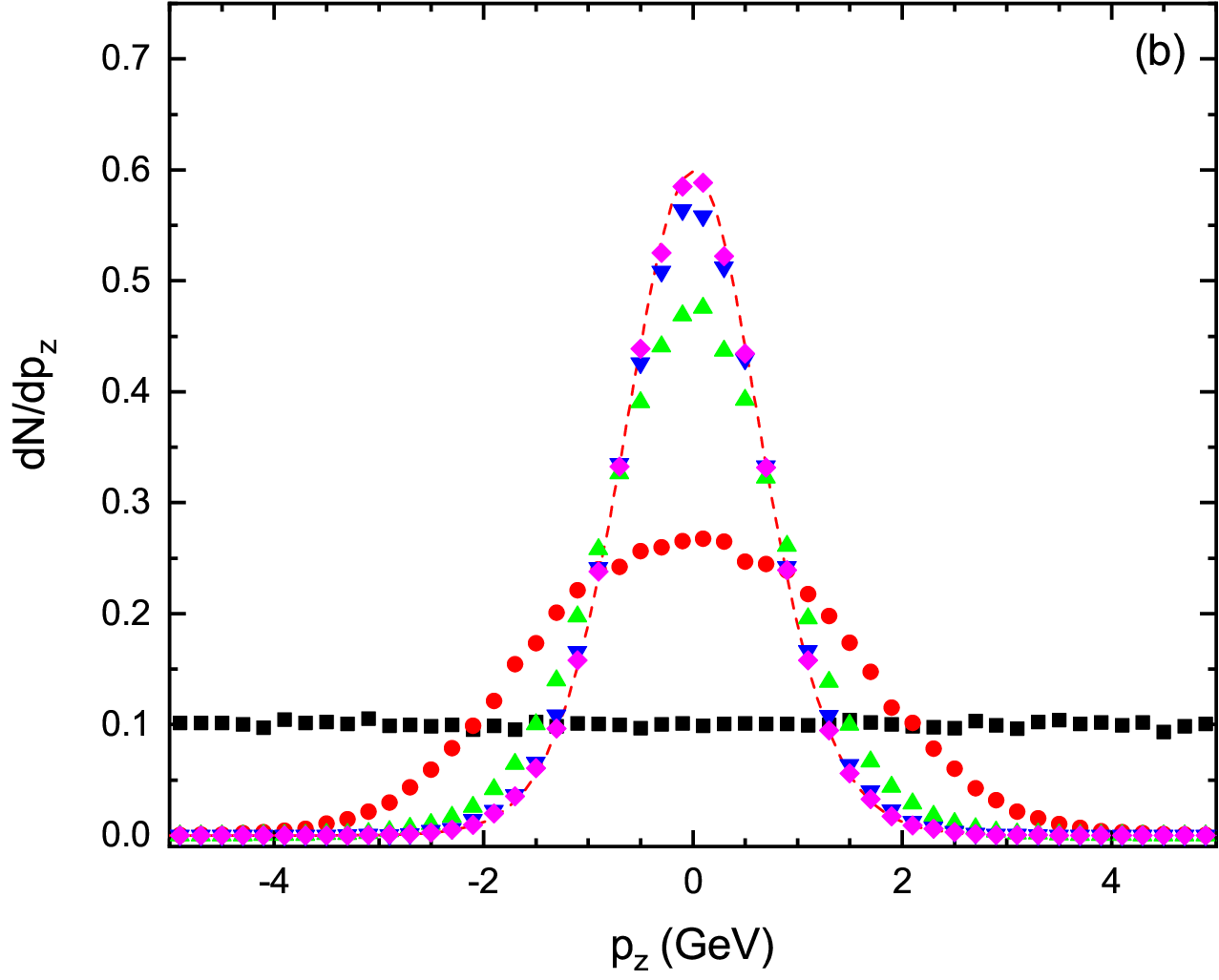}
	\caption{The evolution of charm quark normalized $p_T$ (upper panel) and $p_z$ (lower panel) distributions through Langevin simulation at $1.3 T_c$, by using a constant thermal relaxation rate $\gamma=0.1~\text{fm}^{-1}$. The red dashed lines represent the Boltzmann distributions of charm quarks.}
	\label{fig_Langevin}
\end{figure}

\begin{figure} [htbp]
	\includegraphics[width=0.8\columnwidth]{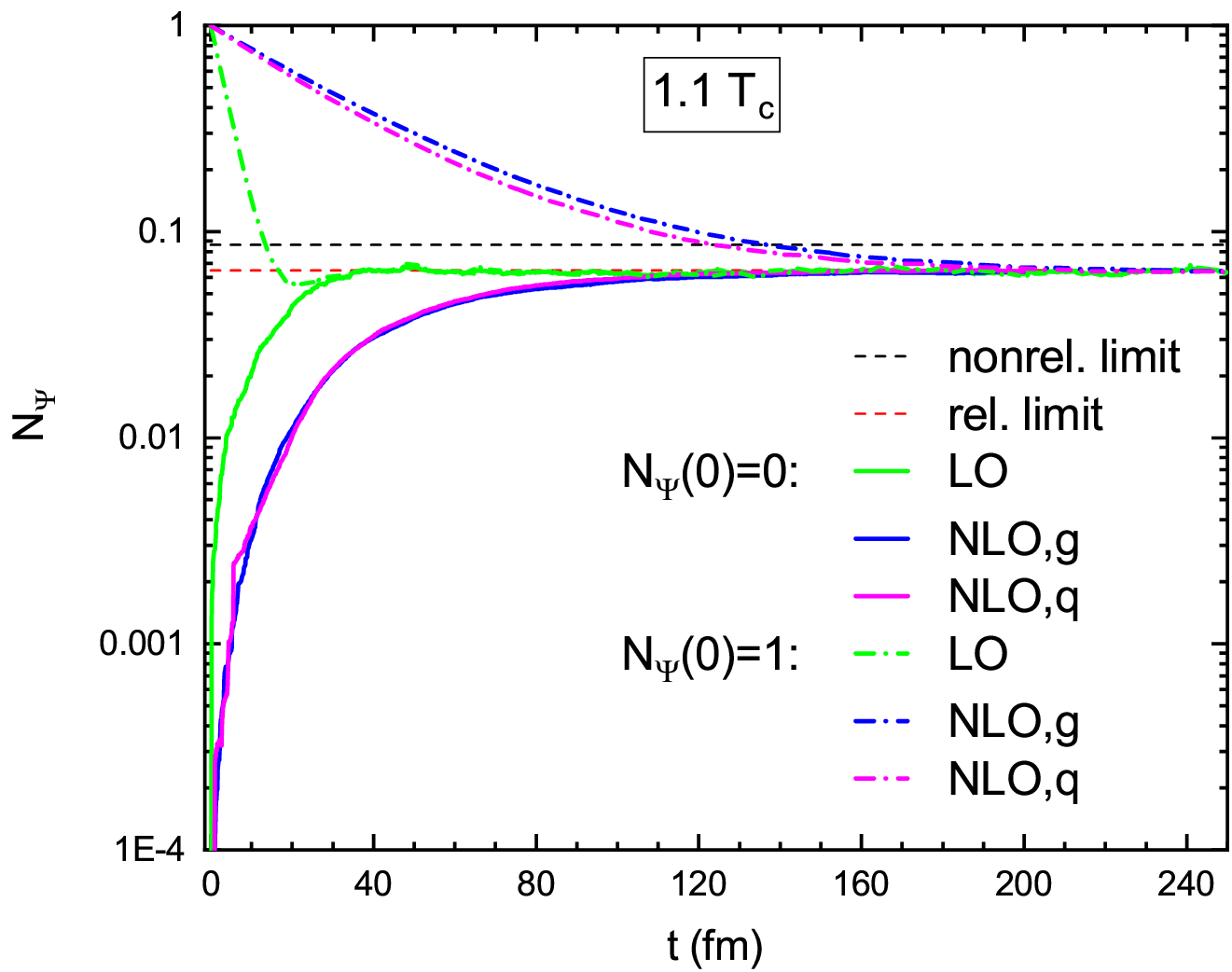}
	\includegraphics[width=0.8\columnwidth]{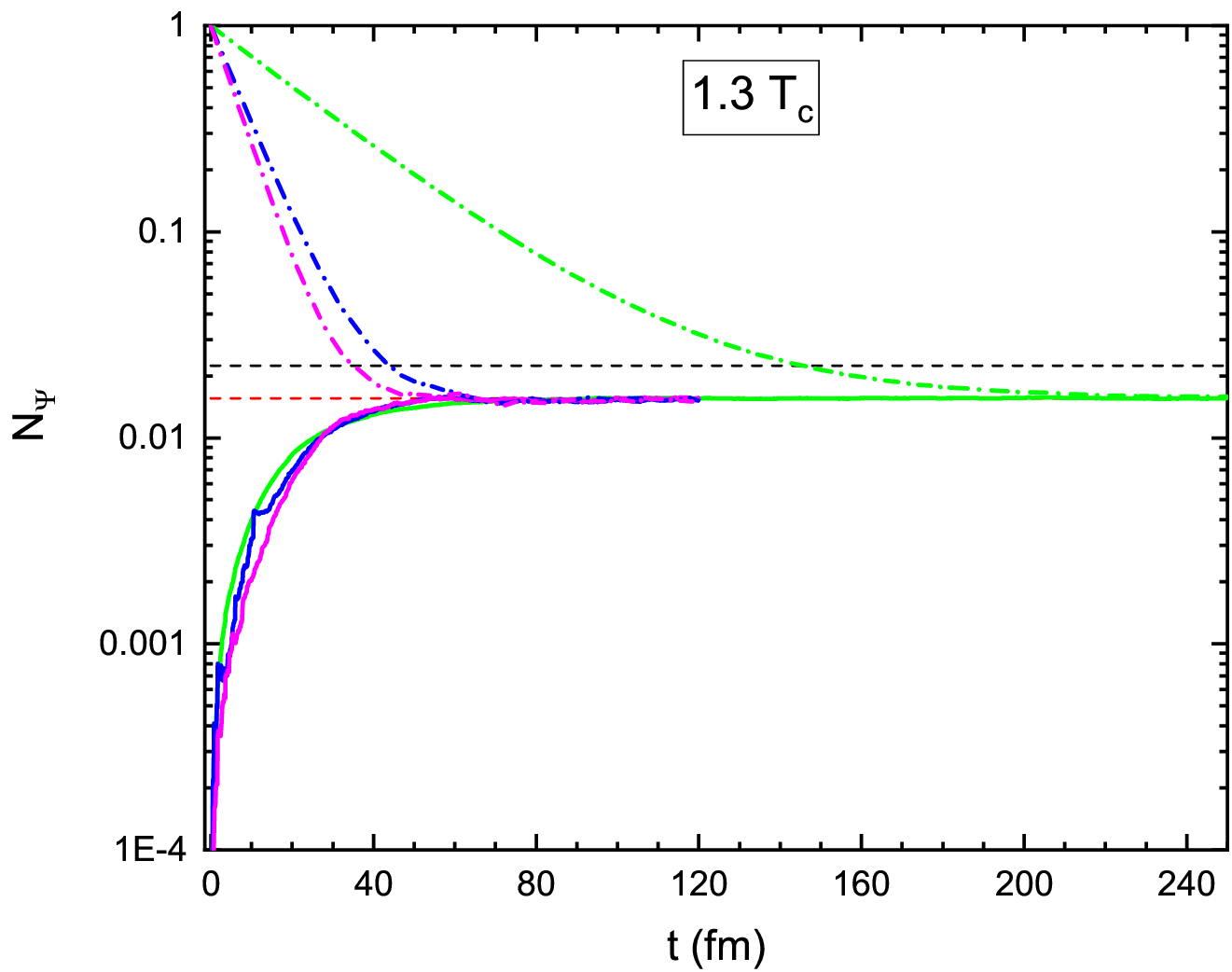}
	\includegraphics[width=0.8\columnwidth]{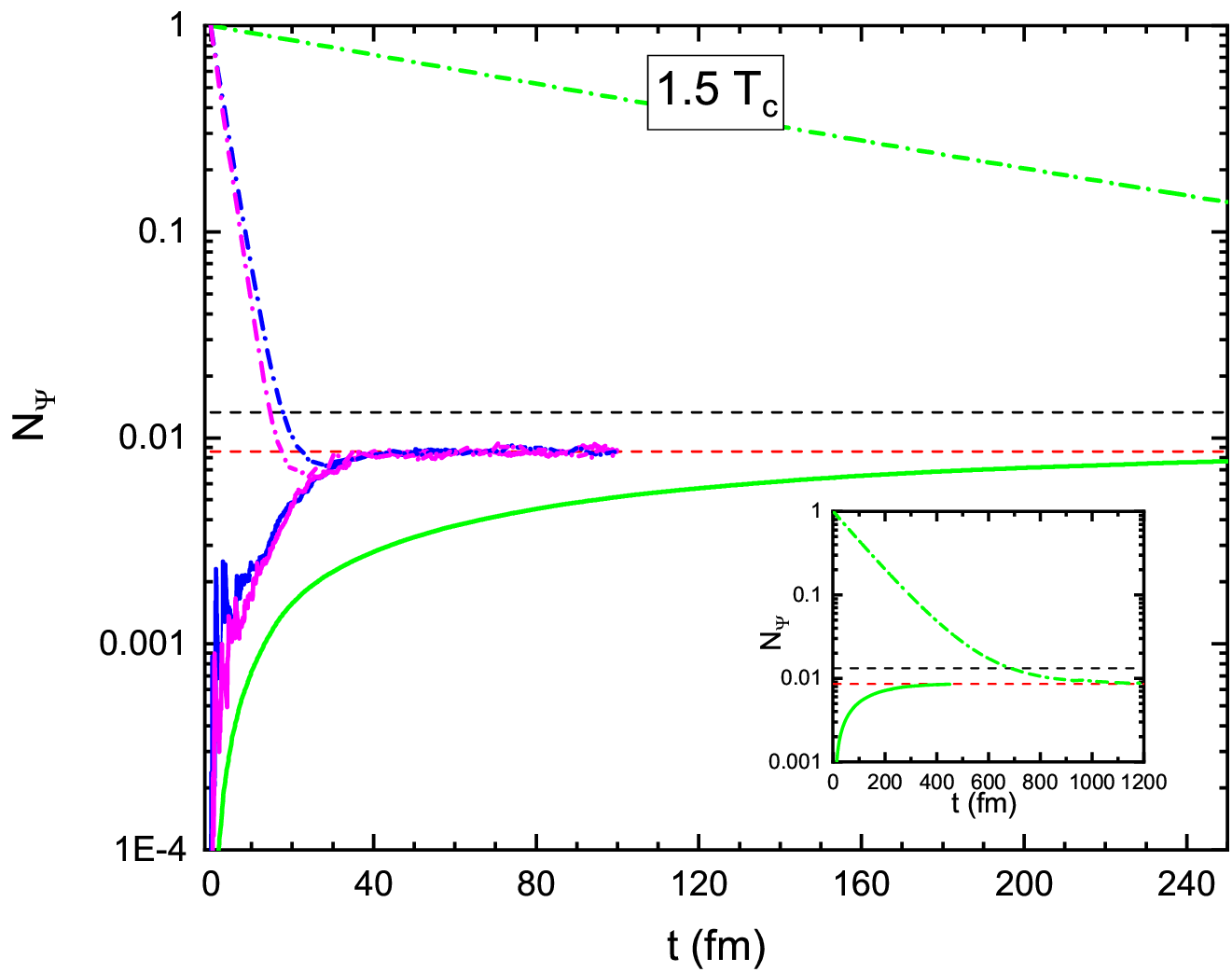}
	\caption{Similar to Fig.~\ref{fig_input_thermalized_Q}; but the transport of bound state is coupled with the single heavy quark Langevin simulations, \ie, regeneration here is computed from transported $c$ and $\bar{c}$. }
	\label{fig_with_Langevin}
\end{figure}

\begin{figure*}[htbp]
	\includegraphics[width=0.4\linewidth]{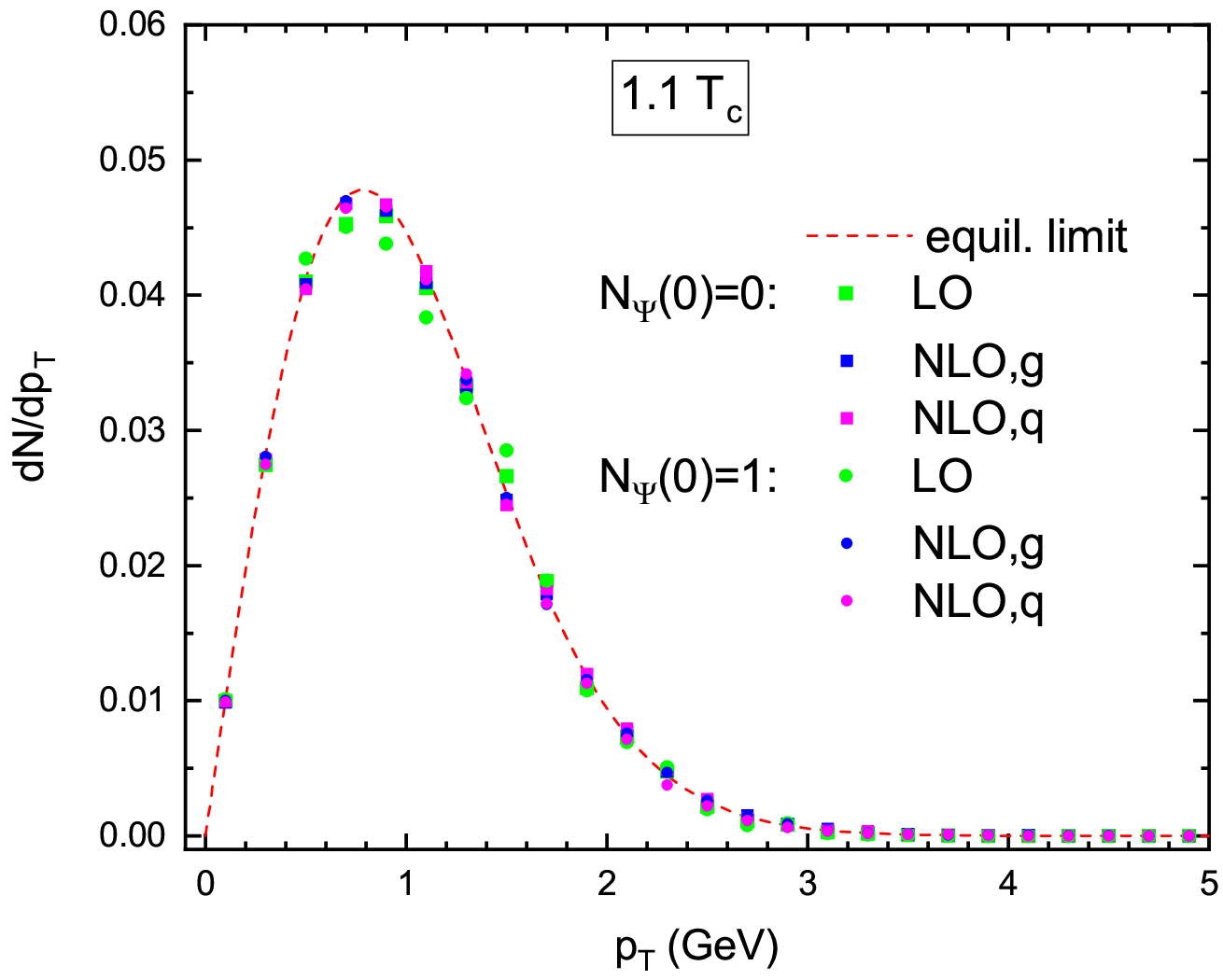}
	\includegraphics[width=0.4\linewidth]{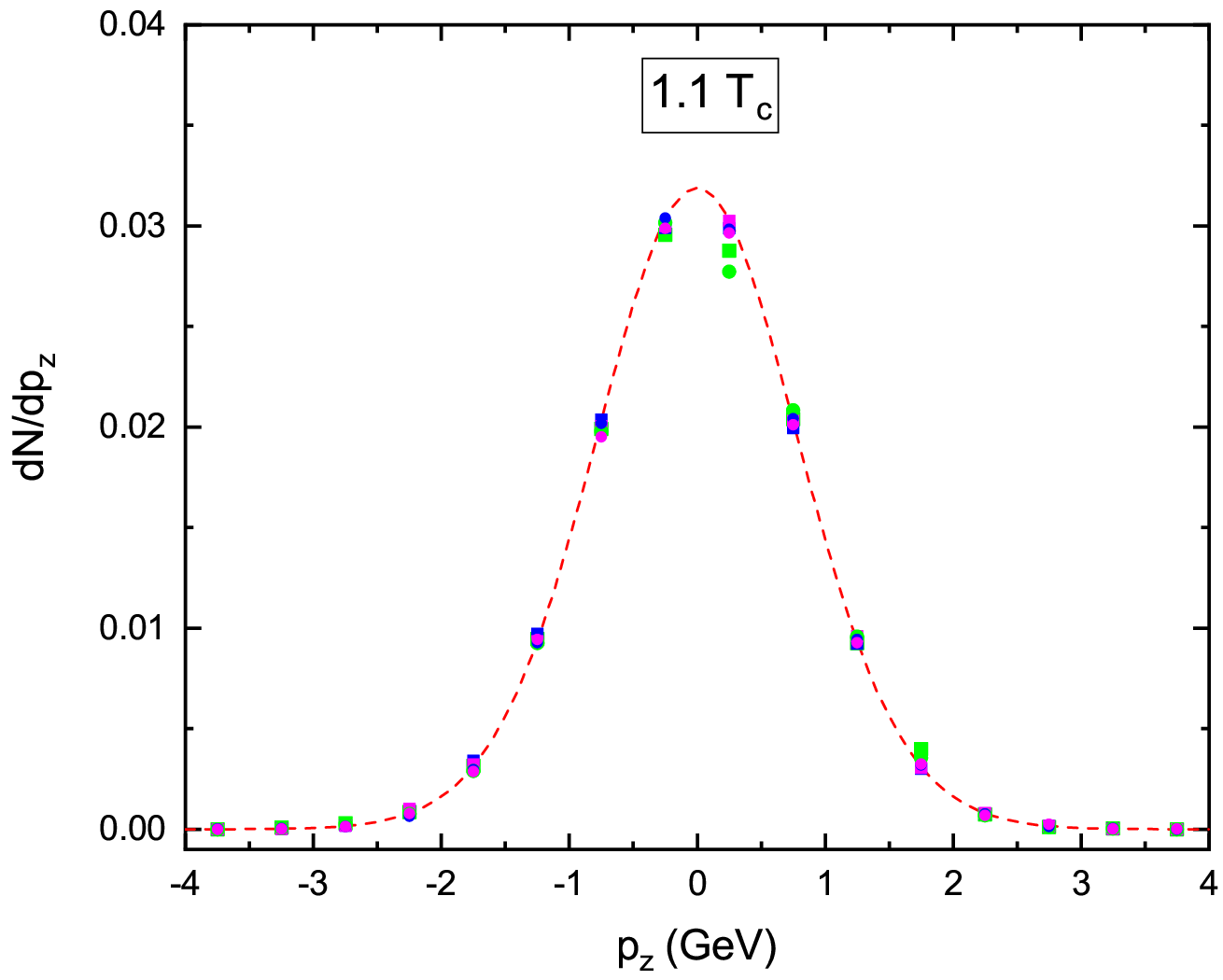}
	\includegraphics[width=0.4\linewidth]{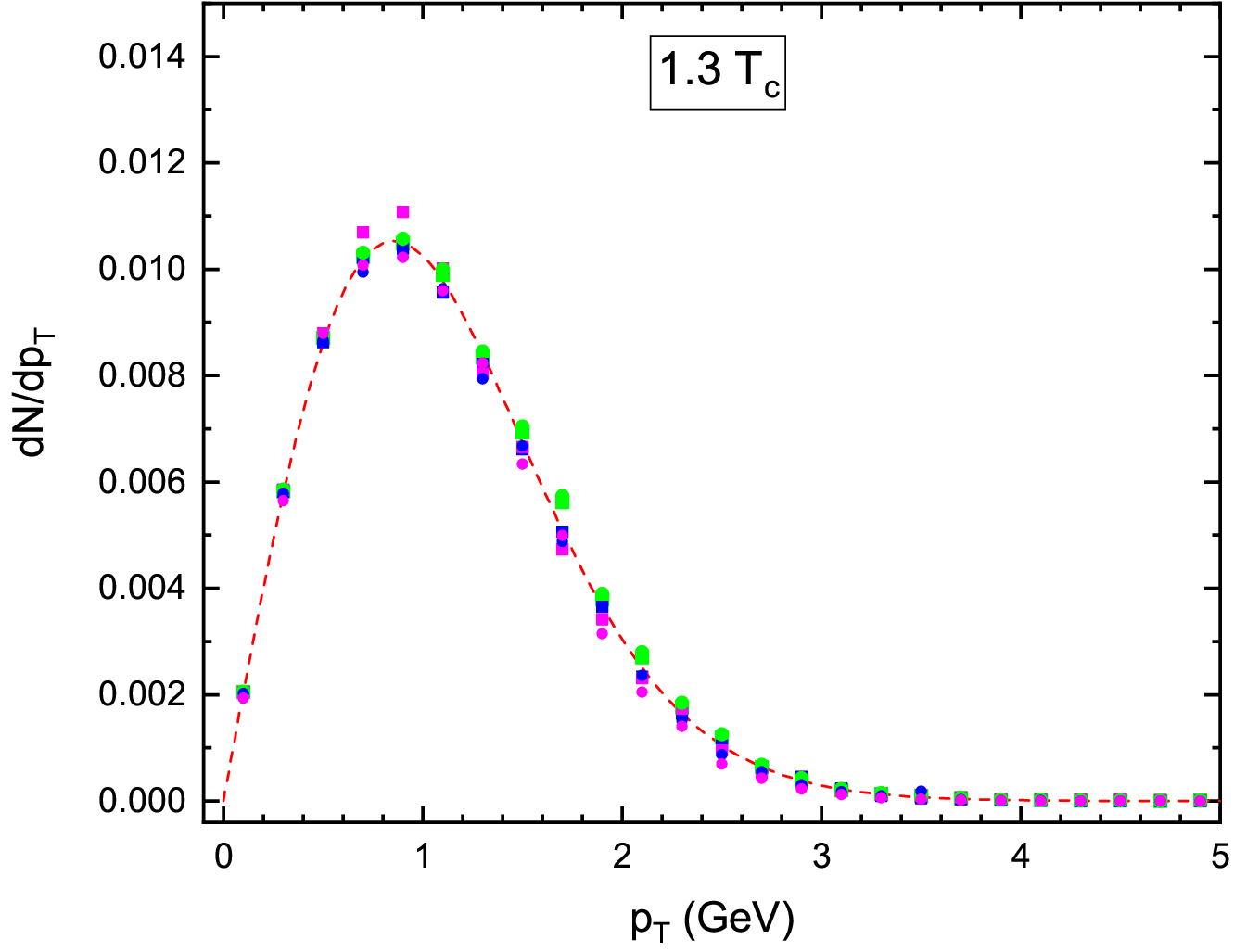}
	\includegraphics[width=0.4\linewidth]{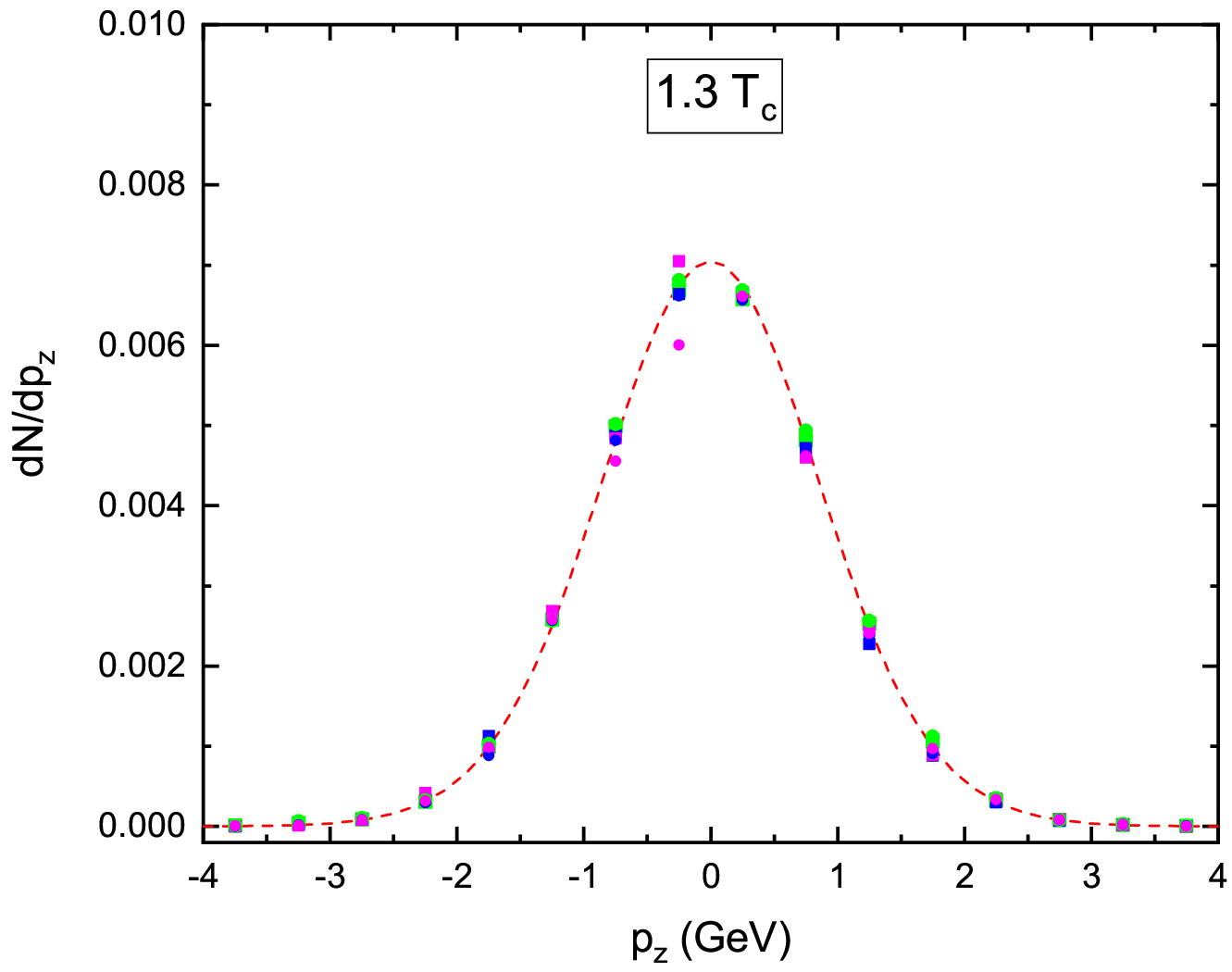}
	\includegraphics[width=0.4\linewidth]{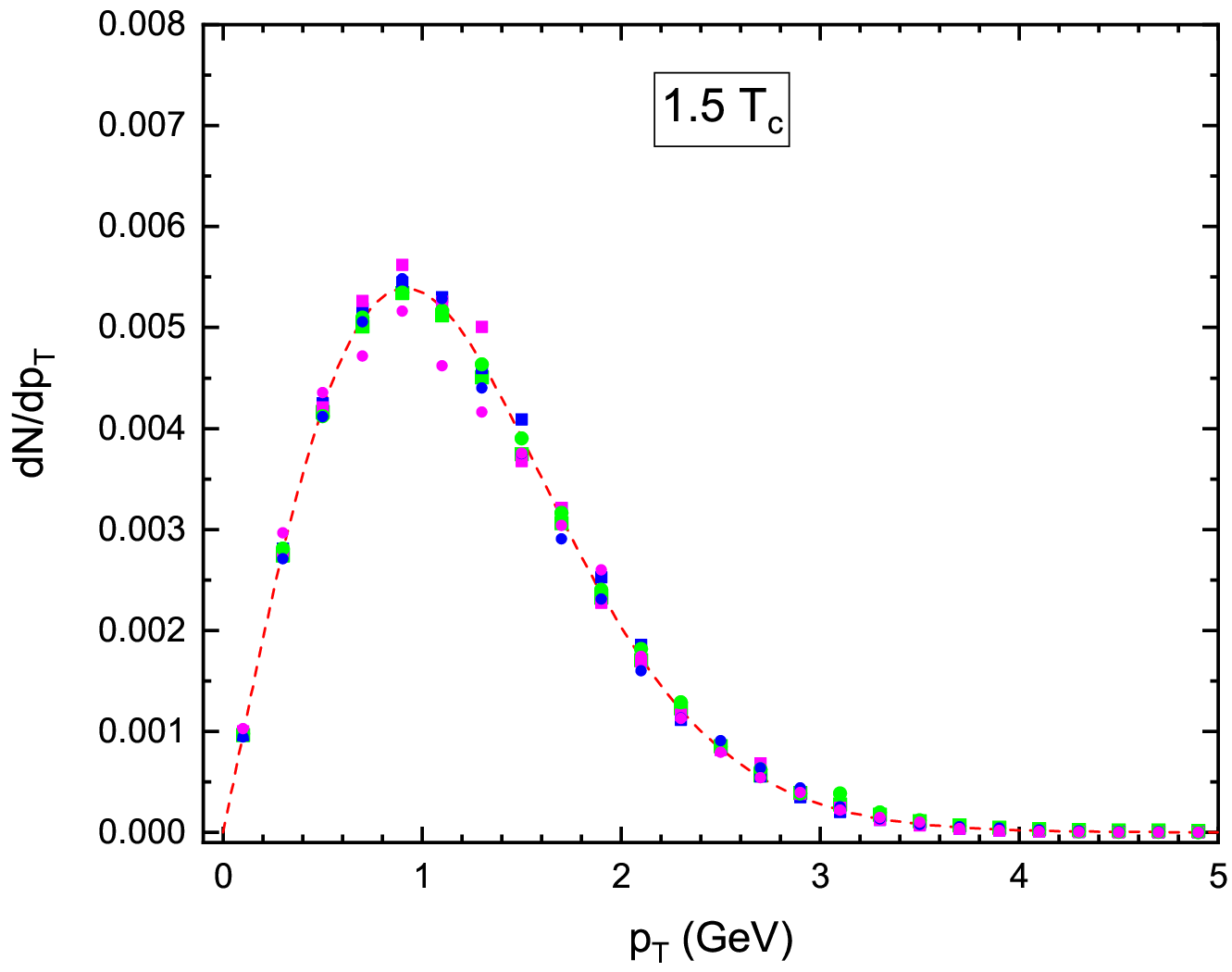}
	\includegraphics[width=0.4\linewidth]{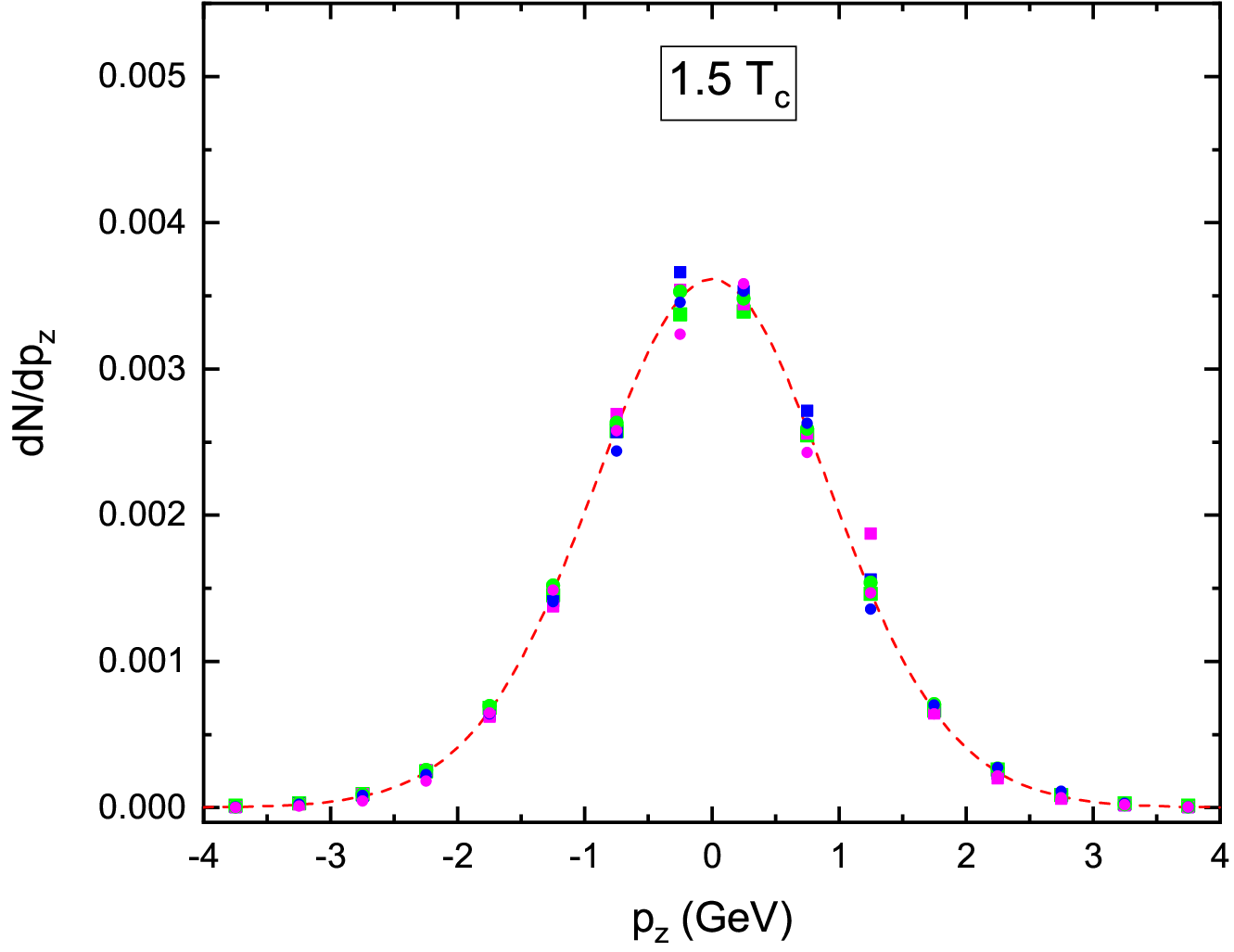}
	\caption{The left and right panels present respectively the $p_T$ and $p_z$ distributions of $J/\psi$ after the relative chemical equilibrium has been reached. The red dashed lines represent the Boltzmann distributions of $J/\psi$ with relativistic dispersion relation at varying temperatures; and the norms correspond to the chemical equilibrium values presented in the horizontal red dashed lines in Figs.~\ref{fig_input_thermalized_Q} and \ref{fig_with_Langevin}.}
	\label{fig_with_Langevin_pT_pz}
\end{figure*}

Now we turn to the simulations of heavy quarkonium equilibration with more realistic non-equilibrium (\ie, transported) single heavy quarks.
In heavy-ion collisions the heavy quarks are produced from the initial hard processes and thus their initial momentum distributions are far from thermalized. Once the QGP forms, the heavy quarks diffuse and thermalize in the medium through rescatterings with thermal partons~\cite{He:2022ywp}. The diffusion of heavy quarks can be simulated by the Langevin equations~\cite{He:2013zua}
\begin{align}\label{Lang_Eq}
	\ud \vec{x} = \frac{\vec{p}}{E_p}\ud t,~~~~~	\ud \vec{p} = -\Gamma(p,T)\vec{p}\ud t + \sqrt{2\ud t D(\vec{p}+\ud\vec{p})}\vec{\rho} ,
\end{align}
where $E_p=\sqrt{m^2+\vec{p}^2}$. For the purpose of illustration, we use a constant (momentum- and temperature-independent) drag coefficient (or thermal relaxation rate), $\Gamma(p,T)=\gamma=0.1~\text{fm}^{-1}$. The momentum diffusion coefficient in Eq.~(\ref{Lang_Eq}) is related to the relaxation rate via the Einstein relation, $D(p)=\gamma E_p T$~\cite{He:2013zua}. The $c$ quark diffusion in the static QGP box is then simulated utilizing the post-point Langevin scheme~\cite{He:2013zua}, with their initial momenta being uniformly distributed in the range $-5~\text{GeV} < p_i < 5~\text{GeV}$, $i=1,2,3$. The time evolution for the $c$ quark $p_T$ and $p_z$ distributions are displayed in Fig.~\ref{fig_Langevin} for fixed temperature 1.3$T_c$. It takes $\sim $30-40 fm for these momentum distributions to approach the Boltzmann equilibrium limit as indicated by the red dashed lines in Fig.~\ref{fig_Langevin}, which can be regarded as the kinetic equilibration (\ie, thermalization) time for heavy quarks.

Then we perform the simulation of heavy quarkonium equilibration by coupling the bound state dissociation and regeneration dynamics (Eqs.~(\ref{diss_simul}), (\ref{gain_LO}) and (\ref{gain_NLO})) with the Langevin diffusion of single heavy quarks (Eq.~(\ref{Lang_Eq})) in a real-time fashion. The results for the evolution of the bound state yield for different reactions are shown in Fig.~\ref{fig_with_Langevin}  at different temperatures.
Comparing to the case when heavy quarks are assumed to be fully thermalized from the beginning, as discussed in Sec.~\ref{subsect:thermal_QQbar}, now it generally takes longer time for the quarkonium $\Psi$ to reach chemical equilibrium. For example at 1.3$T_c$, the equilibration time via the NLO reactions now reaches $\sim 50$ fm with initial condition $N_{\Psi}(0)=0$, which reduces to $\sim 20$ fm if heavy quarks are always thermalized (see the middle panel of Fig.~\ref{fig_input_thermalized_Q}). This is because the regeneration rates are sensitively dependent upon the degree of the heavy quark thermalization - regeneration is generally delayed for non-thermalized heavy quarks. Consequently, quarkonium equilibration must lag behind the kinetic thermalization of heavy quarks which takes 30-40 fm at 1.3$T_c$ as shown in Fig.~\ref{fig_Langevin}.

We also note that now, at a given temperature the equilibration times are almost equal for two different initial conditions, in cases where the reaction rates are sufficiently large (\eg, the LO reaction at 1.1$T_c$, the NLO reactions at 1.3$T_c$ and 1.5$T_c$). This is because, with initial condition $N_{\Psi}(0)=1$, the reaction dynamics in the early stage is dominated by the dissociation while the regeneration is still inefficient (heavy quarks still being far from thermalized). Only after the initial bound state is mostly gone and the heavy quarks thermalize to a large extent, regeneration proceeds efficiently toward the equilibrium limit, making the equilibration time insensitive to the initial conditions. Actually as manifested in the LO reaction at 1.1$T_c$ (dash-dotted green line in upper panel of Fig.~\ref{fig_with_Langevin}) and the NLO reaction at 1.5$T_c$ (dash-dotted purple/blue line in lower panel of Fig.~\ref{fig_with_Langevin}) with $N_{\Psi}(0)=1$, the early time dissociation is so much dominant that the resulting bound state yield can fall below the equilibrium limit. This deficiency relative to the equilibrium yield is then compensated for by the efficient regeneration after the heavy quarks get fully thermalized, so that finally the bound state yield approaches the equilibrium limit.

Finally the quarkonium $p_T$ and $p_z$ distributions after the {\it relative} chemical equilibrium for the quarkonium yields has been reached (Fig.~\ref{fig_with_Langevin}) are shown in Fig.~\ref{fig_with_Langevin_pT_pz}. These distributions ({\it absolute} spectra without artificial shift in norm) turn out to agree well with Boltzmann distributions (red dashed lines) underlying the equilibrium limit in the balance equation Eq.~(\ref{balanceEq}). This implies, as expected, that kinetic equilibrium for the momentum distributions is already embodied in the (relative) chemical equilibrium for the quarkonium yield.

\subsection{Equilibration with total reaction rates}
\label{subsect:equil-total-rates}

\begin{figure} [htbp]
	\includegraphics[width=0.8\columnwidth]{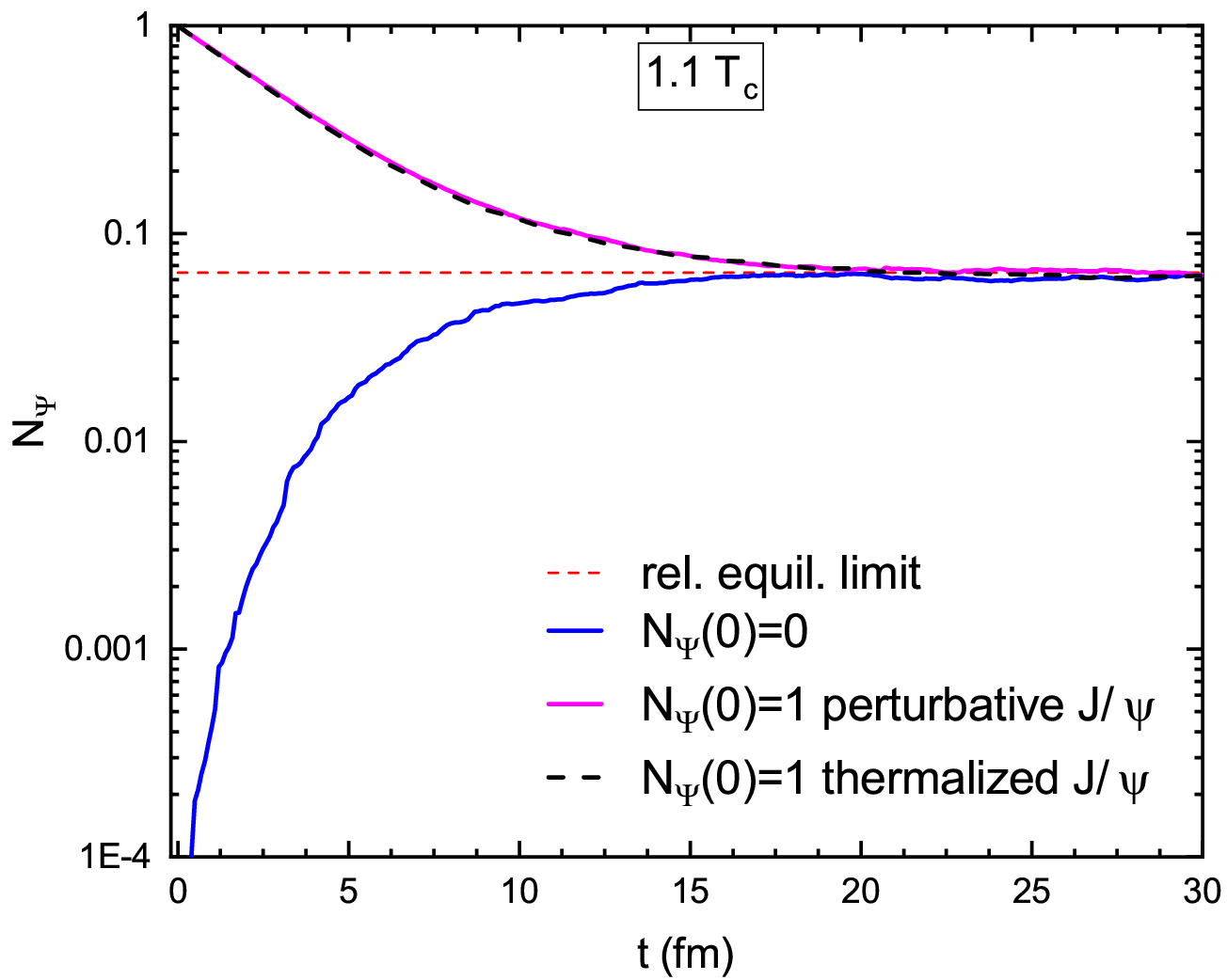}
	\includegraphics[width=0.8\columnwidth]{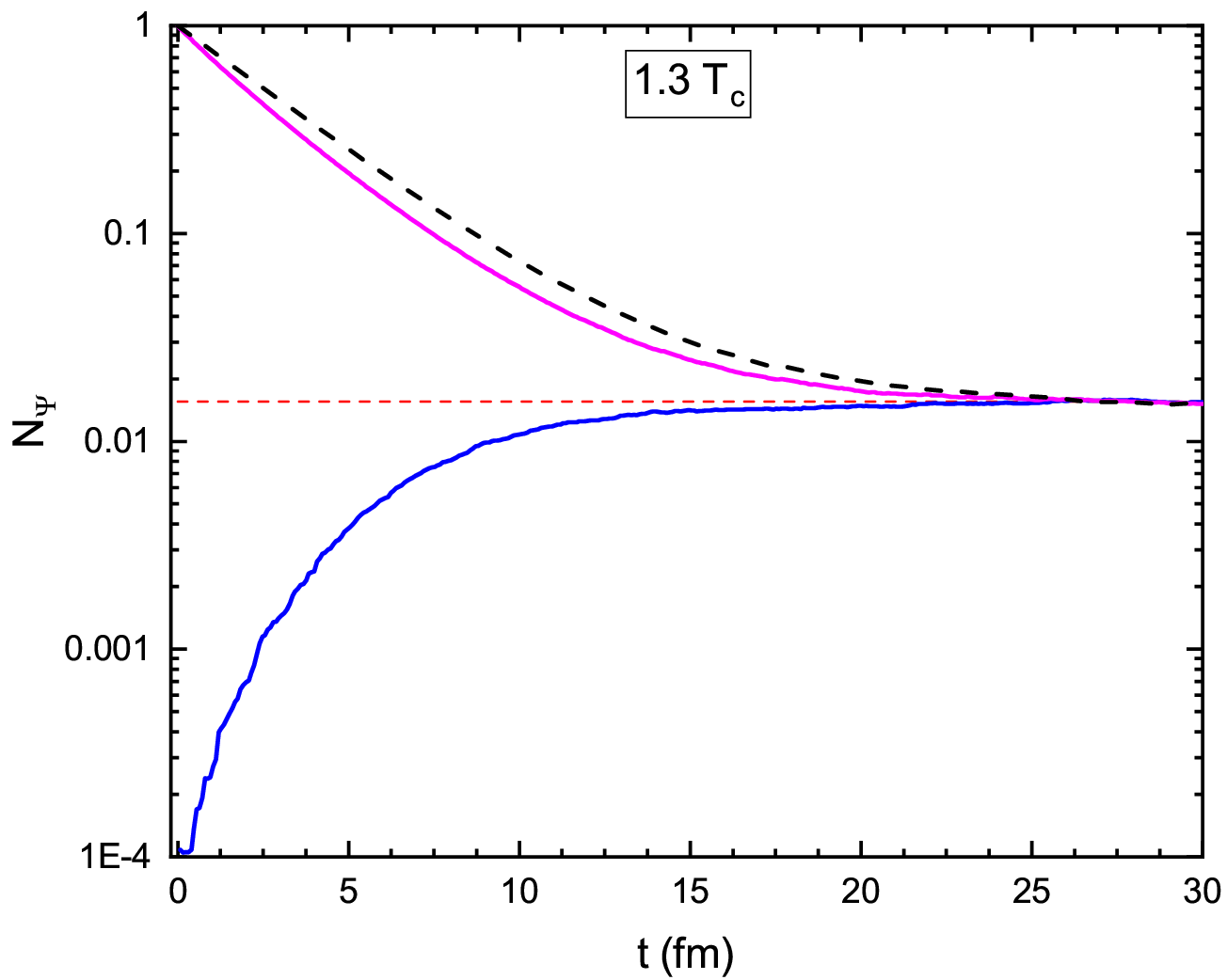}
	\includegraphics[width=0.8\columnwidth]{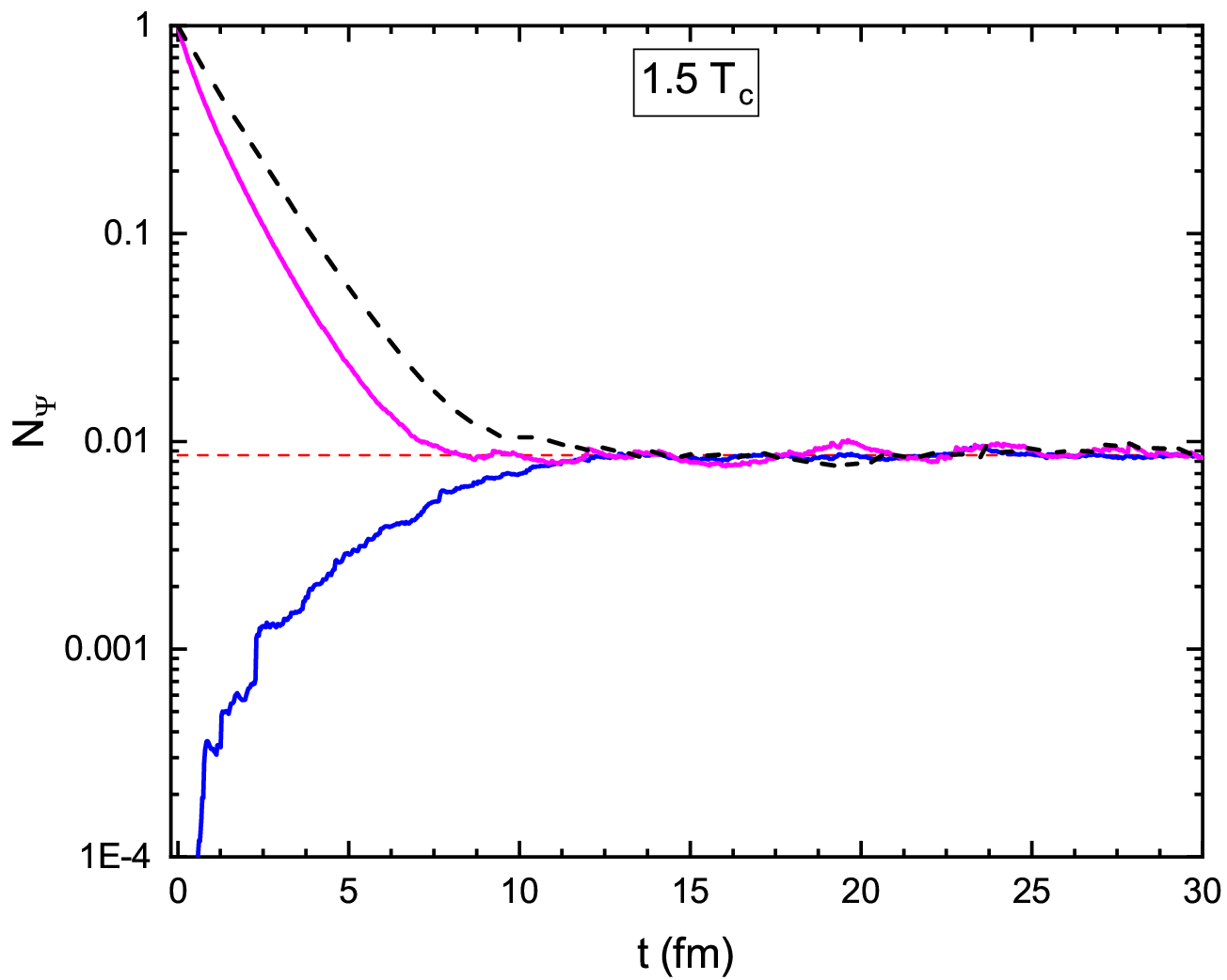}
	\caption{The evolution of $J/\psi$ yield for total reaction rates and transported $c$ and $\bar{c}$ with $\gamma=0.4~{\rm fm^{-1}}$ at 1.1$T_c$ (top), 1.3$T_c$ (middle) and 1.5$T_c$ (bottom). Different initial conditions are compared. The horizontal red dashed lines are the same as shown in Figs.~\ref{fig_input_thermalized_Q} and \ref{fig_with_Langevin}.}
	\label{fig_results-total-rates}
\end{figure}

Having theoretically analyzed and compared the heavy quarkonium equilibration processes for various LO and NLO reactions, we finally conduct a simulation for the equilibration using the total reaction rates (LO and NLO combined) and transported charm quarks with phenomenologically more relevant yet still constant thermal relaxation rate $\gamma=0.4$ $\rm fm^{-1}$~\cite{He:2019vgs,Tang:2023lcn}, in order to make a quantitatively more realistic estimate for the heavy quarkonium equilibration times. For this, the initial momentum distributions for single charm quark and also for the bound state $J/\psi$ (for the initial condition $N_{\Psi}(0)=1$) used in Sec.~\ref{subsect:withLangevin} are also replaced with the more realistic ones from perturbative calculations in $\sqrt{s}=5.02$\,TeV proton-proton collisions~\cite{He:2019vgs,He:2021zej}. The results are shown in Fig.~\ref{fig_results-total-rates} for three temperatures. Now with the total quarkonium reaction rates and larger charm quark thermalization rates, the equilibration time for the $J/\psi$ yield generally comes to the order of 10-15~fm, which is roughly comparable to the lifetime of QGP created in the most central Pb-Pb collisions at the LHC energies. This equilibration time is not very sensitive to the temperatures explored, nor to the choice of initial conditions. For the latter, when there is one bound state initially ($N_{\Psi}(0)=1$), we have compared the results from using two kinds of momentum distributions (perturbative~\cite{He:2021zej} vs. thermalized) for that bound state at the beginning. It turns out that at 1.3$T_c$ and 1.5$T_c$ where the NLO dissociation rates (increasing with momentum, cf. Fig.~\ref{fig_diss_rates}) dominate, the perturbative momentum distribution that entails a larger mean momentum permits a faster equilibration (solid purple curve in Fig.~\ref{fig_results-total-rates}) than the case with thermal momentum distribution (dashed black curve).

\section{Summary}
\label{sect:sum}
In this work we have established a semi-classical transport framework to simulate heavy quarkonium chemical equilibration through regeneration and dissociation dynamics in the QGP, including the leading-order $2\leftrightarrow2$ and next-to-leading order $2\leftrightarrow3$ processes. The scattering amplitudes are taken from our perturbative calculations~\cite{Chen:2017jje,Zhao:2024gxt} using color-electric dipole coupling vertex from QCD multipole expansion. Utilizing the test-particle representations for the heavy quark distribution, the Boltzmann equation describing the time evolution of quarkonium distributions was reduced to tractable iteration schemes for the dissociation of a moving quarkonium in QGP and the regeneration from uncorrelated heavy quark and antiquark pairs.

Numerically, we first simulated the quarkonium (the ground state charmonium $J/\psi$) equilibration in a static and homogeneous QGP box by assuming fully thermalized distributions for heavy quarks participating in recombination. As expected from the rate equations (Eqs.~(\ref{rate_eq_LO}) and (\ref{rate_eq_NLO})), given sufficiently long time the integrated yields of $J/\psi$ can always reach the equilibrium limit as governed by the statistical balance equation (\ref{balanceEq}), regardless of the initial number of bound states. Different speeds toward the equilibrium limit were also compared for various LO and NLO reactions at different temperatures. Next we considered a more realistic case, in which heavy quarks and antiquarks are not thermalized from the beginning but undergo diffusion and gradual thermalization as simulated by Langevin equations. The quarkonium Boltzmann transport is then coupled with the heavy quark diffusion in a real-time fashion. The yields of quarkonia were shown to reach the expected equilibrium limits as well and their final momentum spectra also exhibited good agreement with the analytical Boltzmann distributions, for various LO and NLO reactions. Yet the equilibration time generally becomes significantly longer than the former case, because now the regeneration rates involving non-equilibrium heavy quarks in the early stages of the evolution are suppressed compared to the case with completely thermalized heavy quarks, highlighting the interplay between the open and hidden charm sector in quantifying the heavy quarkonium equilibration. We have also found that the equilibration time is almost independent of the initial number of bound state when the reaction rates are sufficiently large. With total quarkonium reaction rates and more realistic charm quark thermal relaxation rate ($\gamma\sim0.4~{\rm fm^{-1}}$), the equilibration time reduces to the order of 10-15 fm that is comparable to the QGP lifetime in central heavy-ion collisions at the LHC energies. We note that the reaction rates used in the present work are taken from perturbative calculations~\cite{Chen:2017jje,Zhao:2024gxt}, which may be further enhanced by some nonperturbative mechanism~\cite{Tang:2024dkz}. Should larger reaction rates be used, one expects the heavy quarkonium equilibration time to be further reduced, thereby lending a more quantitative understanding of reaction dynamics underlying the statistical hadronization model~\cite{Andronic:2017pug,Andronic:2021erx,Andronic:2025jbp} for charmonium production in heavy-ion collisions.

This work paves the way for realistic phenomenological applications to heavy quarkonium transport. For this, the approach developed here in a static and homogeneous QGP box needs to be extended to incorporate a dynamically evolving medium usually simulated by relativistic hydrodynamics, in which the heavy quarkonium dissociation and regeneration dynamics is to be coupled to the single heavy quark diffusion in a real-time fashion. This would enable us to study the production of heavy quarkonia in heavy-ion collisions in a fully dynamical way, even taking into account the quantum transitions between different bound states~\cite{Zhao:2024gxt}.

\begin{acknowledgments}
This work was supported by the National Natural Science Foundation of China (NSFC) under Grant No. 12475141, the Postgraduate Research \& Practice Innovation Program of Jiangsu Province under KYCX25\_0676, and the Fundamental Research Funds for the Central Universities No. 30925020109.
\end{acknowledgments}

\end{document}